\documentclass[aps,showpacs,showkeys,preprintnumbers,amsmath,amssymb]
{revtex4}
\usepackage{graphicx}
\usepackage{slashed}
\usepackage{mathbbold}
\begin{document}

\title{
Calculation of $1/m^{2}_{b}$ corrections to $\Lambda_{b}\rightarrow\Lambda_{c}$ decay widths in the Bethe-Salpeter equation approach}

\author{ L. Zhang$^{1}$\footnote[1]{
200931220004@mail.bnu.edu.cn}, X.-H. Guo$^{1}$\footnote[2]{ Corresponding author. xhguo@bnu.edu.cn} and M.-H. Weng$^{1}$\footnote[3]{mhweng@mail.bnu.edu.cn
}}
\affiliation{$^{1}$ Jincheng College of Sichuan University, Chengdu 611731, People's Republic of China}
\affiliation{$^{1+}$ College of Nuclear Science and Technology,
Beijing Normal University, Beijing
100875, People's Republic of China }
\affiliation{$^{1\ddag}$ Department of Physics and Electronic Information Engineering, Minjiang University, Fuzhou 350108, People's Republic of China }


\begin{abstract}

The matrix element of the weak transition~$\Lambda_{b}\rightarrow\Lambda_{c}$ can be expressed in terms of six form factors. ~$\Lambda_{Q}~(Q=b,c)$ can be regarded as composed of a heavy quark ~$Q~(Q=b,c)$ and a diquark which is made up of the remaining two light quarks. In this picture, we express these six form factors in terms of Bethe-Salpeter wave functions to second order in the $1/m_{Q}$ expansion. With the kernel containing both the scalar confinement and the one-gluon-exchange terms we calculate the form factors and the decay widths of the semileptonic decay $\Lambda_{b}\rightarrow\Lambda_{c}l\bar{\nu}$ as well as nonleptonic decays $\Lambda_{b}\rightarrow\Lambda_{c}P(V)$ numerically. We also add QCD corrections since they are comparable with $1/m_{Q}$ corrections.

\end{abstract}




\maketitle

\section*{I. INTRODUCTION}

In the past decades more and more experimental data come out with the establishment and running of several high energy accelerators\cite{OPAL}\cite{PDG14}. The focus on heavy quarks has spread from mesons to baryons, partly because of the discovery of the flavor and spin symmetries in QCD, $SU(2)_{f}\times SU(2)_{s}$, in the heavy quark limit and the establishment of the heavy quark effective theory~(HQET)\cite{Isgur1989}-\cite{Georgi1990}. On the experimental side, there have been many new experimental results about $b$-baryons\cite{PDG14}\cite{Akers1996}-\cite{Abe1997}. The lifetime of $\Lambda_{b}$ was given by several experiments\cite{OPAL}. The widths of the nonleptonic decay $\Lambda_{b}\rightarrow \Lambda J/\Psi$\cite{CDF1997} as well as the semileptonic decay $\Lambda_{b}\rightarrow \Lambda_cl^{-}\bar{\nu}$ were measured\cite{DELPHI1995}\cite{DELPHI2004}. Besides, the branching ratios of two-body decays such as $\Lambda_{b}\rightarrow\Lambda_{c}\pi$\cite{CDF2007122002},~$\Sigma^{*}_{c}\rightarrow\Lambda_{c}\pi$ and $\Sigma_c\rightarrow\Lambda_{c}\pi$\cite{CDF2011012003} have been listed in PDG's booklet. CDF has measured decay widths of $\Sigma^{*}_{b}\rightarrow\Lambda_{b}\pi$ and $\Sigma_{b}\rightarrow\Lambda_{b}\pi$ since 2010\cite{CDF2007202001}. All of these measurements help to test theoretical studies for heavy baryons.

Although the heavy quark symmetries can be used to simplify the physical processes where heavy hadrons are involved, in most cases HQET itself can not give the final phenomenological predictions for the decay properties. Hence one still has to adopt nonperturbative QCD models in the end. Among them, there are QCD sum rules, the Bethe-Salpeter~(BS) equation, the chiral perturbation theory, the potential model, the bag model, the instanton model, the relativistic and nonrelativistic quark models, etc.. Applying these models one can calculate weak transition form factors for, for instance, $\Lambda_{b}\rightarrow\Lambda_{c}$, and consequently the semileptonic decay width directly because the lepton pair can be separated from the hadronic weak transition.

The matrix element of $\Lambda_{b}\rightarrow\Lambda_{c}$ can be expressed by six transition form factors because of Lorentz covariance,
\begin{eqnarray}
\langle\Lambda_{c}(v',s')|\bar{c}(\gamma_{\mu}-\gamma_{\mu}\gamma_{5})b|\Lambda_{b}(v,s)\rangle&=&\bar{u}_{\Lambda_{c}}(v',s')[F_{1}(\omega)\gamma_{\mu}+F_{2}(\omega)\nu_{mu}+F_{3}(\omega)\nu'_{\mu}\nonumber\\
&&~~~~~-G_{1}(\omega)\gamma_{\mu}\gamma_{5}-G_{2}(\omega)\nu_{\mu}\gamma_{5}-G_{3}(\omega)\nu'_{\mu}\gamma_{5}]u_{\Lambda_{b}}(v,s),\label{1}
\end{eqnarray}
\noindent where $v'$ and $v$ are the velocities of $\Lambda_{c}$ and $\Lambda_{b}$ respectively, $\omega=v'\cdot v$, $s'$ and $s$ are the spins of $\Lambda_{c}$ and $\Lambda_{b}$ respectively, $u_{\Lambda_c}(v',s')$ and $u_{\Lambda_b}(v,s)$ are the Dirac spinors of $\Lambda_c$ and $\Lambda_b$, respectively, $F_{i}$ and $G_{i} ~(i=1,2,3)$ are the Lorentz scalar form factors. There is only one form factor~(the Isgur-Wise function) remained when we take off all the corrections in the $1/m_{Q}$ expansion. When we take first order corrections into account\cite{Georgi1990},
\begin{eqnarray}
&&F_{1}=G_{1}\Bigg[1+\bigg(\frac{1}{m_{c}}+\frac{1}{m_{b}}\bigg)\frac{\bar{\Lambda}}{1+\omega} \Bigg],\nonumber\\
&&F_{2}=G_{2}=-G_{1}\frac{1}{m_{c}}\frac{\bar{\Lambda}}{1+\omega},\\
&&F_{3}=-G_{3}=-G_{1}\frac{1}{m_{b}}\frac{\bar{\Lambda}}{1+\omega},\nonumber\label{2}
\end{eqnarray}
\noindent where $\bar{\Lambda}$ is a parameter which is defined as the mass difference $m_{\Lambda_{Q}}-m_{Q}$ in the limit $m_{Q}\rightarrow\infty$. There is one independent form factor such as $F_{1}$ in this case.

As a formally exact equation to describe the hadronic bound state, the BS equation is an effective method to deal with nonperturbative QCD effects. With HQET, the BS equation has already been applied to the heavy hadron systems. In the limit $m_{Q}\rightarrow\infty$, we found that the BS equations for the heavy baryons $\Lambda_{Q}$ are greatly simplified. When the quark mass is very heavy compared with the QCD scale, $\Lambda_{QCD}$, the light degrees of freedom in a heavy baryon $\Lambda_{Q}$ become blind to the flavor and spin quantum numbers of the heavy quark because of the $SU(2)_{f}\times SU(2)_{s}$ symmetries. Therefore, the angular momentum and the flavor quantum numbers of the light degrees of freedom become good quantum numbers that can be used to classify heavy baryons, and $\Lambda_{Q}$ corresponds to the state in which the angular momentum of the light degrees of freedom is zero. So it is natural to regard the heavy baryon as composed of one heavy quark and a light diquark. When $1/m_{Q}$ corrections are taken into account, since the isospin of $\Lambda_{Q}$ is zero, the isospin of the light degrees of freedom is also zero, therefore, the spin of the light degrees of freedom should also be zero in order to guarantee that the total wave function of $\Lambda_{Q}$ is antisymmetric. Hence the spin and isospin of the light degrees of freedom are still fixed even when $1/m_{Q}$ corrections are taken into account. Therefore, we still treat $\Lambda_{Q}$
as composed of a heavy quark and a scalar light diquark.  In this picture, we established the BS equations of $\Lambda_{Q}$ to second order in the $1/m_Q$ expansion assuming the kernel contains two parts, a scalar confinement term and a one-gluon-exchange term. In the present work, we will apply the BS equation to calculate the semileptonic and nonleptonic decay widths to second order in the $1/m_Q$ expansion.

The remainder of this paper is organized as follows. In Sec. II we will review the BS equation for $\Lambda_{Q}$ to second order in the $1/m_{Q}$ expansion briefly. In Sec. III we will express the six form factors for $\Lambda_b\rightarrow\Lambda_c$ transition in terms of the BS wave functions of $\Lambda_{Q}$ to second order in the $1/m_{Q}$ expansion and give the numerical results. In Sec. IV we will deduce the numerical results for the semileptonic decay width of $\Lambda_{b}\rightarrow\Lambda_{c}l\bar{\nu}$ and nonleptonic decay widths of $\Lambda_{b}\rightarrow\Lambda_{c}P(V)$~(where $P$ denotes pseudoscalar mesons and $V$ denotes vector mesons
) by applying the numerical results of form factors we get in Sec. III. We will also compare our results with the experimental data. Finally, Sec. V is reserved for summary and discussion.

\section*{II. BS EQUATION FOR ~$\Lambda_{Q}$ TO SECOND ORDER IN THE ~$1/m_{Q}$ EXPANSION}

As we discussed in Introduction, $\Lambda_{Q}$ is regarded as a bound state of a heavy quark and a light scalar diquark. Hence we can define the BS wave function of $\Lambda_{Q}$ as the following\cite{Guo1996}:
\begin{eqnarray}
\chi(x_{1},x_{2},P)=\langle 0|T\psi(x_{1})\varphi(x_{2})|\Lambda_{Q}(P)\rangle,\label{3}
\end{eqnarray}
\noindent where $\psi(x_{1})$ and $\varphi(x_{2})$ are the field operators of the heavy quark at position $x_{1}$ and the light scalar diquark at position $x_{2}$, respectively, $P=m_{\Lambda_{Q}}v$ is the momentum of $\Lambda_{Q}$, and $v$ is its velocity. Let $m_{Q}$ and $m_{D}$ represent the masses of the heavy quark and the light diquark in the baryon,  $\lambda_{1}=\frac{m_{Q}}{m_{Q}+m_{D}}$, $\lambda_{2}=\frac{m_{D}}{m_{Q}+m_{D}}$, and $p$ represent the relative momentum of the two constituents. Then, the BS wave function in momentum space is defined as
\begin{eqnarray}
\chi(x_{1},x_{2},P)=e^{iPX}\int\frac{d^{4}p}{(2\pi)^{4}}e^{ipx}\chi_{P}(p),\label{4}
\end{eqnarray}
\noindent where $X=\lambda_{1}x_{1}+\lambda_{2}x_{2}$ is the coordinate of the center of mass and $x=x_{1}-x_{2}$.

It is straightforward to prove that the BS equation for $\Lambda_{Q}$ has the following form in momentum space\cite{Lurie1968}:
\begin{eqnarray}
\chi_{P}(p)=S_{F}(p_{1})\int\frac{d^{4}q}{(2\pi)^{4}}K(P,p,q)\chi_{P}(q)S_{D}(p_{2}),\label{5}
\end{eqnarray}
\noindent where $p_{1}=\lambda_{1}P+p$, $p_{2}=-\lambda_{2}P+p$ are the momenta of the heavy quark and the light scalar diquark, respectively, $K(P,p,q)$ is the kernel which is defined as the sum of two particle irreducible diagrams, $S_{F}(p_{1})$ and $S_{D}(p_{2})$ are propagators of the heavy quark with momentum $p_{1}$ and the light diquark with momentum $p_{2}$,
\begin{eqnarray}
S_{F}(p_{1})=\frac{i}{\slashed{p}_{1}-m_{Q}+i\varepsilon},\label{6}
\end{eqnarray}
\begin{eqnarray}
S_{D}(p_{2})=\frac{1}{p_{2}^{2}-m_{D}^2+i\varepsilon}.\label{7}
\end{eqnarray}

In order to solve the BS equation for $\Lambda_{Q}$ to second order in the $1/m_{Q}$ expansion, we rewrite Eq.~(\ref{5}) as the following:
\begin{eqnarray}
\chi_{0P}(p)+\frac{1}{m_{Q}}\chi_{1P}(p)+\frac{1}{m_{Q}^{2}}\chi_{2P}(p)&=&\bigg(S_{0F}(p_{1})+\frac{1}{m_{Q}}S_{1F}(p_{1})+\frac{1}{m_{Q}^{2}}S_{2F}(p_{1})\bigg)\int\frac{d^{4}q}{(4\pi)^{4}}\bigg(K_{0}(P,p,q)\nonumber\\
&&+\frac{1}{m_{Q}}K_{1}(P,p,q)+\frac{1}{m_{Q}^{2}}K_{2}(P,p,q)\bigg)\times\Big(\chi_{0P}(q)+\frac{1}{m_{Q}}\chi_{1P}(q)\nonumber\\
&&+\frac{1}{m_{Q}^{2}}\chi_{2P}(q)\Big)S_{D}(p_{2}),\label{8}
\end{eqnarray}
\noindent where we have expanded the BS wave function, the propagator of the heavy quark, and the kernel to $1/m^{2}_{Q}$.

It is easy to show that $S_{D}$ remains unchanged in the $1/m_{Q}$ expansion. Then, by comparing the two sides of Eq.~(\ref{8}) at each order in $1/m_{Q}$, we have the following equations:

\noindent to leading order in the ~$1/m_{Q}$ expansion:
\begin{eqnarray}
\chi_{0P}(p)=S_{0F}(p_{1})\int\frac{d^{4}q}{(2\pi)^4}K_{0}(P,p,q)\chi_{0P}(q)S_{D}(p_{2}),\label{9}
\end{eqnarray}
to first order in the~$1/m_{Q}$ expansion:
\begin{eqnarray}
\chi_{1P}(p)&=&S_{1F}(p_{1})\int\frac{d^{4}q}{(2\pi)^{4}}K_{0}(P,p,q)\chi_{0P}(q)S_{D}(p_{2})+S_{0F}(p_{1})\int\frac{d^{4}q}{(2\pi)^{4}}K_{1}(P,p,q)\chi_{0P}(q)S_{D}(p_{2})\nonumber\\
&&+S_{0F}(p_{1})\int\frac{d^{4}q}{(2\pi)^{4}}K_{0}(P,p,q)\chi_{1P}(q)S_{D}(p_{2}),\label{10}
\end{eqnarray}
to second order in the~$1/m_{Q}$ expansion:
\begin{eqnarray}
\chi_{2P}(p)&=&S_{2F}(p_{1})\int\frac{d^{4}q}{(2\pi)^{4}}K_{0}(P,p,q)\chi_{0P}(q)S_{D}(p_{2})+S_{1F}(p_{1})\int\frac{d^{4}q}{(2\pi)^{4}}K_{1}(P,p,q)\chi_{0P}(q)S_{D}(p_{2})\nonumber\\
&&+S_{1F}(p_{1})\int\frac{d^{4}q}{(2\pi)^{4}}K_{0}(P,p,q)\chi_{1P}(q)S_{D}(p_{2})+S_{0F}(p_{1})\int\frac{d^{4}q}{(2\pi)^{4}}K_{2}(P,p,q)\chi_{0P}(q)S_{D}(p_{2})\nonumber\\
&&+S_{0F}(p_{1})\int\frac{d^{4}q}{(2\pi)^{4}}K_{1}(P,p,q)\chi_{1P}(q)S_{D}(p_{2})+S_{0F}(p_{1})\int\frac{d^{4}q}{(2\pi)^{4}}K_{0}(P,p,q)\chi_{2P}(q)S_{D}(p_{2}).\label{11}
\end{eqnarray}

In our previous work\cite{Zhang2013}, we found that:
\begin{eqnarray}
&&\chi_{0P}(p)=\phi_{0P}(p)u_{\Lambda_{Q}}(v,s),\nonumber\\
&&\chi^{+}_{1,2P}(p)=\frac{1+\slashed{v}}{2}\chi_{1,2P}(p)=\phi^{+}_{1,2P}(p)u_{\Lambda_{Q}}(v,s),\\
&&\chi^{-}_{1,2P}(p)=\frac{1-\slashed{v}}{2}\chi_{1,2P}(p)=\phi^{-}_{1,2P}(p)\slashed{p}_{t}u_{\Lambda_{Q}}(v,s)\nonumber\label{12}
\end{eqnarray}
\noindent where we use the variables $p_{l}=v\cdot p-\lambda_{2}m_{\Lambda_{Q}}$ and $p_{t}=p-(v\cdot p)v$,  $\phi_{0P}(p)$ and $\phi^{\pm}_{1,2P}(p)$ are scalar functions.

As before, we assume the kernel has the following form:
\begin{eqnarray}
&&-iK_{0}=1\otimes1V_{1}+v^{\mu}\otimes(p_{2}+p'_{2})^{\mu}V_{2},\nonumber\\
&&-iK_{1}=1\otimes1V_{3}+\gamma^{\mu}\otimes(p_{2}+p'_{2})^{\mu}V_{4},\\
&&-iK_{2}=1\otimes1V_{5}+\gamma^{\mu}\otimes(p_{2}+p'_{2})^{\mu}V_{6},\nonumber\label{13}
\end{eqnarray}
\noindent where the first parts on the right hand represent the scalar confinement terms while the second parts represent the one-gluon-exchange terms.

\begin{center}
\begin{figure}[htbp]
\includegraphics[height=6cm]{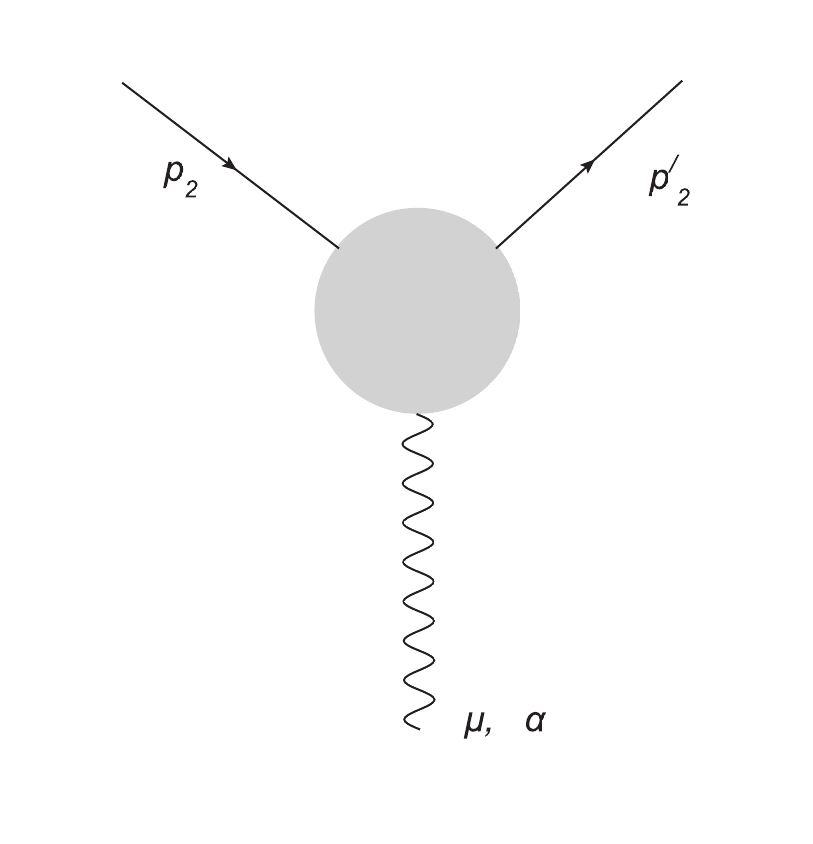}
\caption{~Diquark-gluon-diquark vetex, $\mu$ and $\alpha$ are the Lorentz and color indices of the gluon, respectively.}
\end{figure}
\end{center}

Defining the variable $W^{2}_{P}=\sqrt{p^{2}_{t}+m^{2}_{D}}$ and using the following relation
\begin{eqnarray}
m_{\Lambda_{Q}}=m_{Q}+m_{D}+E_{0}+\frac{1}{m_{Q}}E_{1}+\frac{1}{m^{2}_{Q}}E_{2},\label{14}
\end{eqnarray}
\noindent where $E_{0}$, $E_{1}$, and $E_{2}$ represent the binding energies, we have the BS equations for the scalar functions in Eq.~(\ref{12}) in leading order, first order and second order in the $1/m_{Q}$ expansion:

\begin{eqnarray}
&&\tilde{\phi}_{0P}(p_{t})=-\frac{1}{2(m_{D}+E_{0}-W_{P})(W_{P})}\int\frac{d^{3}q_{t}}{(2\pi)^{3}}[\tilde{V}_{1}-2W_{P}\tilde{V}_{2}]\tilde{\phi}_{0P}(q_{t}),\label{15}\\
&&\tilde{\phi}^{+}_{1P}(p_{t})=0,\label{16}\\
&&\tilde{\phi}^{-}_{1P}(p_{t})=\frac{1}{2}\tilde{\phi}_{0P}(p_{t}),\label{17}\\
&&\tilde{\phi}^{+}_{2P}(p_{t})=\frac{-1}{2(m_{D}+E_{0}-W_{P})W_{P}}\int\frac{d^{3}q_{t}}{(2\pi)^{3}}[\tilde{V}_{1}-2W_{P}\tilde{V}_{2}]\tilde{\phi}^{+}_{2P}(q_{t})\nonumber\\
&&~~~~~~~~~~~~~~+\frac{-1}{2(m_{D}+E_{0}-W_{P})W_{P}}\int\frac{d^{3}q_{t}}{(2\pi)^{3}}(p_{t}\cdot q_{t}-q^{2}_{t})\tilde{V}_{4}\tilde{\phi}^{-}_{1P}(q_{t})\nonumber\\
&&~~~~~~~~~~~~~~+\frac{-1}{2(m_{D}+E_{0}-W_{P})W_{P}}\int\frac{d^{3}q_{t}}{(2\pi)^{3}}[\tilde{V}_{5}-2W_{P}\tilde{V}_{6}]\tilde{\phi}_{0P}(q_{t})\nonumber\\
&&~~~~~~~~~~~~~~+\frac{-1}{4(m_{D}+E_{0}-W_{P})W_{P}}\int\frac{d^{3}q_{t}}{(2\pi)^{3}}[\tilde{V}_{1}-2W_{P}\tilde{V}_{2}]p_{t}\cdot q_{t}\tilde{\phi}^{-}_{1P}(q_{t})\nonumber\\
&&~~~~~~~~~~~~~~+\Big[\frac{p^{2}_{t}}{8(m_{D}+E_{0}-W_{P})W_{P}}+\frac{E_{2}}{(p_{l}+m_{D}+E_{0}+i\varepsilon)^{2}(p^{2}_{l}-W^{2}_{P}+i\varepsilon)}\Big]\nonumber\\
&&~~~~~~~~~~~~~~\times\int\frac{d^{3}q_{t}}{(2\pi)^{3}}[\tilde{V}_{1}-2W_{P}\tilde{V}_{2}]\tilde{\phi}_{0P}(q_{t})\nonumber\\
&&~~~~~~~~~~~~~~+\frac{-1}{4(m_{D}+E_{0}-W_{P})W_{P}}\int\frac{d^{3}q_{t}}{(2\pi)^{3}}[p_{t}\cdot q_{t}-p^{2}_{t}]\tilde{V}_{4}\tilde{\phi}_{0P}(q_{t}),\label{18}\\
&&\tilde{\phi}^{-}_{2P}(p)=\frac{1}{4W_{P}}\int\frac{d^{3}q_{t}}{(2\pi)^{3}}[\tilde{V}_{1}-2W_{P}\tilde{V}_{2}]\frac{p_{t}\cdot q_{t}}{p_{t}\cdot p_{t}}\tilde{\phi}^{-}_{1P}(q_{t})\nonumber\\
&&~~~~~~~~~~~~~~+\frac{1}{4W_{P}}\int\frac{d^{3}q_{t}}{(2\pi)^{3}}\bigg(1+\frac{p_{t}\cdot q_{t}}{p_{t}\cdot p_{t}}\bigg)\tilde{V}_{4}\tilde{\phi}_{0P}(q)\nonumber\\
&&~~~~~~~~~~~~~~+\frac{1}{8W_{P}}\int\frac{d^{3}q_{t}}{(2\pi)^{3}}[\tilde{V}_{1}-2W_{P}\tilde{V}_{2}]\tilde{\phi}_{0P}(q),\label{19}
\end{eqnarray}
\noindent where we have defined $\tilde{\phi}(p_{t})\equiv\int\frac{dp_{l}}{2\pi}\phi(p)$ and applied the covariant instantaneous approximation in the kernel,~$\tilde{V_i}=V_i|_{p_l=q_l}~(i=1,\ldots6)$.
The numerical results of these scalar functions were given in our previous work\cite{Zhang2013}, and in that paper we discussed why Eqs.~(\ref{16}) and (\ref{17}) hold.

\section*{III. $\Lambda_{b}\rightarrow\Lambda_{c}$ FORM FACTORS TO $1/m^{2}_{Q}$ }
In this section we will express the six form factors for the $\Lambda_{b}\rightarrow\Lambda_{c}$ weak transition in terms of the BS wave functions to second order in the $1/m_{Q}$ expansion and give their numerical results.
On the grounds of Lorentz invariance, the matrix element for $\Lambda_{b}\rightarrow\Lambda_{c}$ can be expressed as in Eq.~(\ref{1}).

On the other hand, the matrix element for $\Lambda_{b}\rightarrow\Lambda_{c}$ can be related to the BS wave functions of $\Lambda_{b}$ and $\Lambda_{c}$ as the following:
\begin{eqnarray}
\langle\Lambda_{c}(v',s')|\bar{c}(\gamma_{\mu}-\gamma_{\mu}\gamma_{5})b|\Lambda_{b}(v,s)\rangle=\int\frac{d^{4}p}{(2\pi)^{4}}\bar{\chi}_{P'}(p')(\gamma_{\mu}-\gamma_{\mu}\gamma_{5})\chi_{P}(p)S^{-1}_{D}(p_{2}),\label{20}
\end{eqnarray}
\noindent where $P'~(P)$ is the momentum of $\Lambda_{b}~(\Lambda_{c})$ and $p~(p')$ is the relative momentum defined in the BS wave function of $\Lambda_{b}(v,s)~(\Lambda_{c}(v',s'))$.
As we did before, we can express the BS wave functions of $\Lambda_{b}$ and $\Lambda_{c}$ in the terms of the scalar functions $\phi_{0P}(p)$ and $\phi^{\pm}_{1,2P}(p)$ to second order in the $1/m_{Q}$ expansion
\begin{eqnarray}
\bar{\chi}_{P'}(p')&=&\bar{\chi}_{0P'}(p')+\frac{1}{m_{c}}\bar{\chi}_{1P'}(p')+\frac{1}{m^{2}_{c}}\bar{\chi}_{2P'}(p')\nonumber\\
&=&\bar{u}_{\Lambda_{c}}(v',s')\bigg[\phi_{0P'}(p')+\frac{1}{m_{c}}\bigg(\phi^{+}_{1P'}(p')+\slashed{p}'_{t}\phi^{-}_{1P'}(p')\bigg)+\frac{1}{m^{2}_{c}}\bigg(\phi^{+}_{2P'}(p')+\slashed{p}'_{t}\phi^{-}_{2P'}(p')\bigg)\bigg],\label{21}\\
\chi_{P}(p)&=&\chi_{0P}(p)+\frac{1}{m_{b}}\chi_{1P}(p)+\frac{1}{m^{2}_{b}}\chi_{2P}(p)\nonumber\\
&=&\bigg[\phi_{0P}(p)+\frac{1}{m_{b}}\bigg(\phi^{+}_{1P}(p)+\slashed{p}_{t}\phi^{-}_{1P}(p)\bigg)+\frac{1}{m^{2}_{b}}\bigg(\phi^{+}_{2P}(p)+\slashed{p}_{t}\phi^{-}_{2P}(p)\bigg)\bigg]u_{\Lambda_{b}}(v,s).\label{22}
\end{eqnarray}
Substituting the above twe equations into Eq.~(\ref{20}) and comparing with Eq.~(\ref{1}) we can obtain the form factors in terms of the BS wave functions. To simplify our results, we define
\begin{eqnarray}
&&\int\frac{d^{4}p}{(2\pi)^{4}}\phi_{\alpha P'}(\beta p')\phi_{P}(p)(p^{2}_{l}-W^{2}_{p})=F_{(\alpha P',\beta P)},\label{23}\\
&&\int\frac{d^{4}p}{(2\pi)^{4}}\phi_{\alpha P'}(p')p_{t\nu}\phi_{\beta P}(p)(p^{2}_{l}-W^{2}_{p})=f_{1(\alpha P',\beta P)}v_{\nu}+f_{2(\alpha P',\beta P)}v'_{\nu},\label{24}\\
&&\int\frac{d^{4}p}{(2\pi)^{4}}\phi_{\alpha P'}(p')p'_{t\nu}\phi_{\beta P}(p)(p^{2}_{l}-W^{2}_{p})=f_{3(\alpha P',\beta P)}v_{\nu}+f_{4(\alpha P',\beta P)}v'_{\nu},\label{25}\\
&&\int\frac{d^{4}p}{(2\pi)^{4}}\phi_{\alpha P'}(p')p'_{t\mu}p_{t\nu}\phi_{\beta P}(p)(p^{2}_{l}-W^{2}_{p})=f_{5(\alpha P',\beta P)}g_{\mu\nu}+f_{6(\alpha P',\beta P)}v'_{\mu}v_{\nu}+f_{7(\alpha P',\beta P)}v_{\mu}v'_{\nu},\label{26}
\end{eqnarray}
\noindent where $F_{(\alpha P',\beta P)}$ and $f_{i(\alpha P',\beta P)}~(i=1,\ldots7)$ are functions of $\omega$, and $\alpha,~\beta=0,1,2$. With the aid of $p_{t\nu}v^{\nu}=p_{t}\cdot v=0$,~$p'_{t\nu}v'^{\nu}=p'_{t}\cdot v'=0$ and $v^{2}=v'^{2}=1$, it is easy to see that
\begin{eqnarray}
&&f_{1(\alpha P',\beta P)}=-\omega f_{2(\alpha P',\beta P)},\label{27}\\
&&f_{2(\alpha P',\beta P)}=\frac{1}{1-\omega^{2}}\int\frac{d^{4}p}{(2\pi)^{4}}\phi_{\alpha P'}(p')p_{t}\cdot v'\phi_{\beta P}(p)(p^{2}_{l}-W^{2}_{P}),\label{28}\\
&&f_{3(\alpha P',\beta P)}=\frac{1}{1-\omega^{2}}\int\frac{d^{4}p}{(2\pi)^{4}}\phi_{\alpha P'}(p')p'_{t}\cdot v\phi_{\beta P}(p)(p^{2}_{l}-W^{2}_{P}),\label{29}\\
&&f_{4(\alpha P',\beta P)}=-\omega f_{3(\alpha P',\beta P)},\label{30}\\
&&f_{5(\alpha P',\beta P)}=\frac{1}{3}\int\frac{d^{4}p}{(2\pi)^{4}}\phi_{\alpha P'}(p')p'_{t}\cdot p_{t}\phi_{P}(\beta p)(p^{2}_{l}-W^{2}_{P}),\label{31}\\
&&f_{6(\alpha P',\beta P)}=0,\label{32}\\
&&f_{7(\alpha P',\beta P)}=-\frac{1}{\omega}f_{5(\alpha P',\beta P)}.\label{33}
\end{eqnarray}
Then, we can express the six form factors, $F_{i},~G_{i}~(i=1,2,3)$, as follows:
\begin{eqnarray}
&&F_{1}(\omega)\nonumber\\
&&=-i\bigg[F_{(0P',0P)}+\frac{1}{m_{b}}\bigg(f_{1(0P',1P^{-})}-f_{2(0P',1P^{-})}\bigg)+\frac{1}{m^{2}_{b}}\bigg(F_{(0P',2P^{+})}+f_{1(0P',2P^{-})}-f_{2(0P',2P^{-})}\bigg)\nonumber\\
&&+\frac{1}{m_{c}}\bigg(-f_{3(1P'^{-},0P)}+f_{4(1P'^{-},0P)}\bigg)+\frac{1}{m^{2}_{c}}\bigg(F_{(2P'^{+},0P)}-f_{3(2P'^{-},0P)}+f_{4(2P'^{-},0P)}\bigg)\nonumber\\
&&+\frac{1}{m_{b}m_{c}}\bigg(-2f_{5(1P'^{-},1P'^{-})}-(1+2\omega)f_{7(1P'^{-},1P'^{-})}\bigg)\bigg],\label{34}
\end{eqnarray}
\begin{eqnarray}
&&F_{2}(\omega)=-i\bigg[\frac{2}{m_{c}}f_{3(1P'^{-},0P)}+\frac{2}{m^{2}_{c}}f_{3(2P'^{-},0P)}+\frac{2}{m_{b}m_{c}}\bigg(f_{3(1P'^{-},1P^{+})}+f_{7(1P'^{-},1P^{-})}\bigg)\bigg],\label{35}\\
&&F_{3}(\omega)=-i\bigg[\frac{2}{m_{b}}f_{2(0P',1P^{-})}+\frac{2}{m^{2}_{b}}f_{2(0P',2P^{-})}+\frac{2}{m_{b}m_{c}}\bigg(f_{2(1P'^{+},1P^{-})}+f_{7(1P'^{-},1P^{-})}\bigg)\bigg],\label{36}
\end{eqnarray}
\begin{eqnarray}
&&G_{1}(\omega)\nonumber\\
&&=-i\bigg[F_{(0P',0P)}+\frac{1}{m_{b}}\bigg(f_{1(0P',1P^{-})}+f_{2(0P',1P^{-})}\bigg)+\frac{1}{m^{2}_{b}}\bigg(F_{(0P',2P^{+})}+f_{1(0P',2P^{-})}+f_{2(0P',2P^{-})}\bigg)\nonumber\\
&&+\frac{1}{m_{c}}\bigg(f_{3(1P'^{-},0P)}+f_{4(1P'^{-},0P)}\bigg)+\frac{1}{m^{2}_{c}}\bigg(F_{(2P'^{+},0P)}+f_{3(2P'^{-},0P)}+f_{4(2P'^{-},0P)}\bigg)\nonumber\\
&&+\frac{1}{m_{b}m_{c}}\bigg(2f_{5(1P'^{-},1P'^{-})}-(1-2\omega)f_{7(1P'^{-},1P'^{-})}\bigg)\bigg],\label{37}
\end{eqnarray}
\begin{eqnarray}
&&G_{2}(\omega)=-i\bigg[\frac{2}{m_{c}}f_{3(1P'^{-},0P)}+\frac{2}{m^{2}_{c}}f_{3(2P'^{-},0P)}+\frac{2}{m_{b}m_{c}}\bigg(f_{3(1P'^{-},1P^{+})}+f_{7(1P'^{-},1P^{-})}\bigg)\bigg],\label{38}\\
&&G_{3}(\omega)=i\bigg[\frac{2}{m_{b}}f_{2(0P',1P^{-})}+\frac{2}{m^{2}_{b}}f_{2(0P',2P^{-})}+\frac{2}{m_{b}m_{c}}\bigg(f_{2(1P'^{+},1P^{-})}-f_{7(1P'^{-},1P^{-})}\bigg)\bigg],\label{39}
\end{eqnarray}
\noindent where the superscripts "$\pm$" of $P$ or $P'$ correspond to $\phi^{(\pm)}_{1,2P}$ or $\phi^{(\pm)}_{1,2P'}$.

We assume that in the weak transition of $\Lambda_{b}\rightarrow\Lambda_{c}$ the diquark behaves as a spectator, so the four momenta of the diquarks in the initial and final baryons are the same, that is,
\begin{eqnarray}
p_{2}=p'_{2}.\label{40}
\end{eqnarray}
Substituting $p_{2}=p-\lambda_{2}m_{\Lambda_{Q}}v$,~$p=p_{t}+(v\cdot p)v$ and $v\cdot p=p_{l}+\lambda_{2}m_{\Lambda_{Q}}$ into Eq.~(\ref{40}), we get
\begin{eqnarray}
p_{t}+p_{l}v=p'_{t}+p'_{l}v'.\label{41}
\end{eqnarray}
Defining ~$v'_{t}=v'-(v\cdot v')v=v'-\omega v$ and $\theta$ be the angular between $p_{t}$ and $v'_{t}$, we have
\begin{eqnarray}
&&p_{t}\cdot v'_{t}=-|p_{t}||v'_{t}|\cos\theta=-|p_{t}|\sqrt{\omega^{2}-1}\cos\theta,\label{42}\\
&&p'_{l}=-|p_{t}|\sqrt{\omega^{2}-1}\cos\theta+p_{l}\omega,\label{43}\\
&&p'_{t}\cdot v=(1-\omega^{2})p_{l}+\omega|p_{t}|\sqrt{\omega^{2}-1}\cos\theta,\label{44}\\
&&p'_{t}\cdot p_{t}=|p_{t}|^{2}+|p_{t}|^{2}(\omega^{2}-1)\cos^{2}\theta-\omega p_{l}|p_{t}|\sqrt{\omega^{2-1}}\cos\theta,\label{45}\\
&&|p'_{t}|^{2}=|p_{t}|^{2}+|p_{t}|^{2}(\omega^{2}-1)\cos^{2}\theta+p^{2}_{l}(\omega^{2}-1)-2\omega p_{l}|p_{t}|\sqrt{\omega^{2}-1}\cos\theta,\label{46}\\
&&(p'_{l})^{2}-W^{2}_{P'}=p^{2}_{l}-W^{2}_{P}.\label{47}
\end{eqnarray}
Then all $p'_{l}$ and $p'_{t}$ in $F$ and $f_{i}~(i=1-7)$ can be replaced by $p_{l}$ and $p_{t}$. Take $F_{(0P',0P)}$ as an example:
\begin{eqnarray}
&&F_{(0P',0P)}=\int\frac{d^{4}p}{(2\pi)^{4}}\phi_{0P'}(p')\phi_{0P}(p)(p^{2}_{l}-W^{2}_{P})\nonumber\\
&&~~~~~~~~~~~=\int\frac{d^{4}p}{(2\pi)^{4}}\frac{-i}{(p'_{l}+m_{D}+E_{0}+i\varepsilon)(p'^{2}_{l}-W^{2}_{P'}+i\varepsilon)}\int\frac{d^{4}k}{(2\pi)^{4}}\big[V'_{1}(k,p'_{t})+(p'_{l}+k_{l})V'_{2}(k,p'_{t})\big]\phi_{0P'}(k)\nonumber\\
&&~~~~~~~~~~~~~\times\frac{-i}{(p_{l}+m_{D}+E_{0}+i\varepsilon)(p^{2}_{l}-W^{2}_{P}+i\varepsilon)}\int\frac{d^{4}q}{(2\pi)^{4}}\big[V_{1}(q,p_{t})+(p_{l}+q_{l})V_{2}(q,p_{t})\big]\phi_{0P}(q)(p^{2}_{l}-W^{2}_{P})\nonumber\\
&&~~~~~~~~~~~=\int\frac{d^{4}p}{(2\pi)^{4}}\frac{-1}{(p'_{l}+m_{D}+E_{0}+i\varepsilon)(p_{l}+m_{D}+E_{0}+i\varepsilon)(p^{2}_{l}-W^{2}_{P}+i\varepsilon)}\int\frac{d^{4}k}{(2\pi)^{4}}\big[V'_{1}(k,p'_{t})\nonumber\\
&&~~~~~~~~~~~~~+(p'_{l}+k_{l})V'_{2}(k,p'_{t})\big]\phi_{0P'}(k)\int\frac{d^{4}q}{(2\pi)^{4}}\big[V_{1}(q,p_{t})+(p_{l}+q_{l})V_{2}(q,p_{t})\big]\phi_{0P}(q)\nonumber\\
&&~~~~~~~~~~~=\int\frac{d^4p}{(2\pi)^4}\frac{-1}{(-|p_{t}|\sqrt{\omega^{2}-1}\cos\theta+p_{l}\omega+m_{D}+E_{0}+i\varepsilon)(p_{l}+m_{D}+E_{0}+i\varepsilon)(p^{2}_{l}-W^{2}_{P}+i\varepsilon)}\nonumber\\
&&~~~~~~~~~~~~~\times\int\frac{d^{4}k}{(2\pi)^{4}}\big[V'_{1}(k,p'_{t})+(-|p_{t}|\sqrt{\omega^{2}-1}\cos\theta+p_{l}\omega+k_{l})V'_{2}(k,p'_{t})\big]\phi_{0P'}(k)\int\frac{d^{4}q}{(2\pi)^{4}}\big[V_{1}(q,p_{t})\nonumber\\
&&~~~~~~~~~~~~~+(p_{l}+q_{l})V_{2}(q,p_{t})\big]\phi_{0P}(q)|_{p'_{t}=\sqrt{|p_{t}|^{2}+|p_{t}|^{2}(\omega^{2}-1)\cos^{2}\theta+p^{2}_{l}(\omega^{2}-1)-2\omega p_{l}|p_{t}|\sqrt{\omega^{2}-1}\cos\theta}}\nonumber\\
&&~~~~~~~~~~~=\frac{(2\pi i)}{2\pi}\int\frac{d^3p_t}{(2\pi)^3}\frac{-1}{(-|p_{t}|\sqrt{\omega^{2}-1}\cos\theta-W_{P}\omega+m_{D}+E_{0})(-W_{P}+m_{D}+E_{0})(-2W_{P})}\int\frac{d^3k_t}{(2\pi)^3}[\tilde{V}'_1(k_t,p'_t)\nonumber\\
&&~~~~~~~~~~~~+2(-|p_{t}|\sqrt{\omega^{2}-1}\cos\theta-W_{P}\omega)\tilde{V}'_2(k_t,p'_t)]\int\frac{dk_l}{2\pi}\phi_{0P'}(k)\int\frac{d^3q_t}{(2\pi)^3}[\tilde{V}_1(q_t,p_t)-2W_p\tilde{V}_2(q_t,p_t)]\nonumber\\
&&~~~~~~~~~~~~\times\int\frac{dq_l}{2\pi}\phi_{0P}(q_l)|_{p'_{t}=\sqrt{|p_{t}|^{2}+|p_{t}|^{2}(\omega^{2}-1)\cos^{2}\theta+W^{2}_{P}(\omega^{2}-1)+2\omega W_{P}|p_{t}|\sqrt{\omega^{2}-1}\cos\theta}}\nonumber\\
&&~~~~~~~~~~~=\int\frac{d^{3}p_{t}}{(2\pi)^{3}}\frac{-i}{(-|p_{t}|\sqrt{\omega^{2}-1}\cos\theta-W_{P}\omega+m_{D}+E_{0})}\int\frac{d^{3}k_{t}}{(2\pi)^{3}}\big[\tilde{V}'_{1}(k_{t},p'_{t})+2(-|p_{t}|\sqrt{\omega^{2}-1}\cos\theta\nonumber\\
&&~~~~~~~~~~~~~-W_{P}\omega)\tilde{V}'_{2}(k_{t},p'_{t})\big]\tilde{\phi}_{0P'}(k_{t})\frac{1}{(-W_P+m_D+E_0)(-2W_P)}\int\frac{d^3q_t}{(2\pi)^3}[\tilde{V}_1(q_t,p_t)-2W_p\tilde{V}_2(q_t,p_t)]\nonumber\\
&&~~~~~~~~~~~~~\times\tilde{\phi}_{0P}(q_t)|_{p'_{t}=\sqrt{|p_{t}|^{2}+|p_{t}|^{2}(\omega^{2}-1)\cos^{2}\theta+W^{2}_{P}(\omega^{2}-1)+2\omega W_{P}|p_{t}|\sqrt{\omega^{2}-1}\cos\theta}}\nonumber\\
&&~~~~~~~~~~~=\int\frac{d^{3}p_{t}}{(2\pi)^{3}}\frac{-i}{(-|p_{t}|\sqrt{\omega^{2}-1}\cos\theta-W_{P}\omega+m_{D}+E_{0})}\int\frac{d^{3}k_{t}}{(2\pi)^{3}}\big[\tilde{V}'_{1}(k_{t},p'_{t})+2(-|p_{t}|\sqrt{\omega^{2}-1}\cos\theta\nonumber\\
&&~~~~~~~~~~~~~-W_{P}\omega)\tilde{V}'_{2}(k_{t},p'_{t})\big]\tilde{\phi}_{0P'}(k_{t})\tilde{\phi}_{0P}(p_{t})|_{p'_{t}=\sqrt{|p_{t}|^{2}+|p_{t}|^{2}(\omega^{2}-1)\cos^{2}\theta+W^{2}_{P}(\omega^{2}-1)+2\omega W_{P}|p_{t}|\sqrt{\omega^{2}-1}\cos\theta}}\nonumber\\
&&~~~~~~~~~~~\equiv\int\frac{d^{3}p_{t}}{(2\pi)^{3}}g_{0P'}(p_{t},\cos\theta)\tilde{\phi}_{0P}(p_{t}),\label{48}
\end{eqnarray}
\noindent where we have done a contour integration with $p_l=-W_P+i\varepsilon$ as the pole and defined $\tilde{\phi}(q_t)\equiv\int\frac{dq_l}{2\pi}\phi(q)$, and
\begin{eqnarray}
g_{0P'}(p_{t},\cos\theta)&&\equiv\frac{-i}{(-|p_{t}|\sqrt{\omega^{2}-1}\cos\theta-W_{P}\omega+m_{D}+E_{0})}\int\frac{d^{3}k_{t}}{(2\pi)^{3}}\big[\tilde{V}'_{1}(k_{t},p'_{t})+2(-|p_{t}|\sqrt{\omega^{2}-1}\cos\theta\-W_{P}\omega)\nonumber\\
&&\times\tilde{V}'_{2}(k_{t},p'_{t})\big]\tilde{\phi}_{0P'}(k_{t})|_{p'_{t}=\sqrt{|p_{t}|^{2}+|p_{t}|^{2}(\omega^{2}-1)\cos^{2}\theta+W^{2}_{P}(\omega^{2}-1)+2\omega W_{P}|p_{t}|\sqrt{\omega^{2}-1}\cos\theta}}.\label{49}
\end{eqnarray}
In the same way we can also get,
\begin{eqnarray}
&&F_{(0P',2P^{+})}=\int\frac{d^{3}p_{t}}{(2\pi)^{3}}g_{0P'}(p_{t},\cos\theta)\tilde{\phi}^{+}_{2P}(p_{t}),\label{50}\\
&&F_{(2P'^{+},0P)}=\int\frac{d^{3}p_{t}}{(2\pi)^{3}}g_{2P'^{+}}(p_{t},\cos\theta)\tilde{\phi}_{0P}(p_{t}),\label{51}\\
&&f_{1(0P',1P^{-})}=-\omega f_{2(0P',1P^{-})},\label{52}\\
&&f_{1(0P',2P^{-})}=-\omega f_{2(0P',2P^{-})},\label{53}\\
&&f_{2(0P',1P^{-})}=\frac{1}{2(1-\omega^{2})}\int\frac{d^{3}p_{t}}{(2\pi)^{3}}g_{0P'}(p_{t},\cos\theta)\big(-|p_{t}|\sqrt{\omega^{2}-1}\cos\theta\big)\tilde{\phi}_{0P}(p_{t}),\label{54}\\
&&f_{2(0P',2P^{-})}=\frac{1}{(1-\omega^{2})}\int\frac{d^{3}p_{t}}{(2\pi)^{3}}g_{0P'}(p_{t},\cos\theta)\big(-|p_{t}|\sqrt{\omega^{2}-1}\cos\theta\big)\tilde{\phi}^{-}_{2P}(p_{t}),\label{55}\\
&&f_{3(1P'^{-},0P)}=\frac{1}{2(1-\omega^{2})}\int\frac{d^{3}p_{t}}{(2\pi)^{3}}g_{0P'}(p_{t},\cos\theta)\big[W_{P}(\omega^{2}-1)+|p_{t}|\sqrt{\omega^{2}-1}\cos\theta\big]\tilde{\phi}_{0P}(p_{t}),\label{56}\\
&&f_{3(2P'^{-},0P)}=\frac{1}{(1-\omega^{2})}\int\frac{d^{3}p_{t}}{(2\pi)^{3}}g_{2P'^{-}}(p_{t},\cos\theta)\big[W_{P}(\omega^{2}-1)+\omega|p_{t}|\sqrt{\omega^{2}-1}\cos\theta\big]\tilde{\phi}_{0P}(p_{t}),\label{57}\\
&&f_{4(1P'^{-},0P)}=-\omega f_{3(1P'^{-},0P)},\label{58}\\
&&f_{4(2P'^{-},0P)}=-\omega f_{3(2P'^{-},0P)},\label{59}\\
&&f_{5(1P'^{-},1P^{-})}=\frac{1}{12}\int\frac{d^{3}p_{t}}{(2\pi)^{3}}g_{0P'}(p_{t},\cos\theta)\big[|p_{t}|^{2}+|p_{t}|^{2}(\omega^{2}-1)\cos\theta+W_{P}|p_{t}|\omega\sqrt{\omega^{2}-1}\cos\theta\big]\tilde{\phi}_{0P}(p_{t}),\label{60}\\
&&f_{7(1P'^{-},1P^{-})}=-\frac{1}{\omega}f_{5(1P'^{-},1P^{-})},\label{61}
\end{eqnarray}
\noindent where we have defined
\begin{eqnarray}
&&g_{2P'^{+}}(p_{t},\cos\theta)\nonumber\\ &\equiv&\frac{-i}{(-|p_{t}|\sqrt{\omega^{2}-1}\cos\theta-W_{P}\omega+m_{D}+E_{0})}\bigg\{\int\frac{d^{3}k_{t}}{(2\pi)^{3}}\big[\tilde{V}'_{1}(k_{t},p'_{t})+2(-|p_{t}|\sqrt{\omega^{2}-1}\cos\theta-W_{P}\omega)\tilde{V}'_{2}(k_{t},p'_{t})\big]\tilde{\phi}^{+}_{2P'}(k_{t})\nonumber\\
&&+\frac{1}{2}\int\frac{d^{3}k_{t}}{(2\pi)^{3}}\big(2p'_{t}\cdot k_{t}-k^{2}_{t}-p'^{2}_{t}\big)\tilde{V}'_{4}(k_{t},p'_{t})\tilde{\phi}_{0P'}(k_{t})+\int\frac{d^{3}k_{t}}{(2\pi)^{3}}\big[\tilde{V}'_{5}(k_{t},p'_{t})+2(-|p_{t}|\sqrt{\omega^{2}-1}\cos\theta-W_{P}\omega)\tilde{V}'_{6}(k_{t},p'_{t})\big]\nonumber\\
&&~\times\tilde{\phi}_{0P'}(k_{t})+\frac{1}{4}\int\frac{d^{3}k_{t}}{(2\pi)^{3}}\big[\tilde{V}'_{1}(k_{t},p'_{t})+2(-|p_{t}|\sqrt{\omega^{2}-1}\cos\theta-W_{P}\omega)\tilde{V}'_{2}(k_{t},p'_{t})\big]p'_{t}\cdot k_{t}\tilde{\phi}_{0P'}(k_{t})\nonumber\\
&&+\big(\frac{-p'^{2}_{t}}{4}+\frac{-E_{2}}{(-|p_{t}|\sqrt{\omega^{2}-1}\cos\theta-W_{P}\omega+m_{D}+E_{0})}\big)\int\frac{d^{3}k_{t}}{(2\pi)^{3}}\big[\tilde{V}'_{1}(k_{t},p'_{t})+2(-|p_{t}|\sqrt{\omega^{2}-1}\cos\theta-W_{P}\omega)\tilde{V}'_{2}(k_{t},p'_{t})\big]\nonumber\\
&&~\times\tilde{\phi}_{0P'}(k_{t})\bigg\}|_{p'_{t}=\sqrt{|p_{t}|^{2}+|p_{t}|^{2}(\omega^{2}-1)\cos^{2}\theta+W^{2}_{P}(\omega^{2}-1)+2\omega W_{P}|p_{t}|\sqrt{\omega^{2}-1}\cos\theta}},\label{62}\\
&&g_{2P'^{-}}(p_{t},\cos\theta)\nonumber\\
&\equiv&i\frac{1}{4W_{P}}\Bigg[\int\frac{d^{3}k_{t}}{(2\pi)^{3}}[\tilde{V}'_{1}(k_{t},p'_{t})+2(-|p_{t}|\sqrt{\omega^{2}-1}\cos\theta-W_{P}\omega)\tilde{V}'_{2}(k_{t},p'_{t})]\frac{p'_{t}\cdot k_{t}}{p'_{t}\cdot p'_{t}}\phi^{-}_{1P'}(k_{t})\nonumber\\
&&+\int\frac{d^{3}k_{t}}{(2\pi)^{3}}\bigg(1+\frac{p'_{t}\cdot k_{t}}{p'_{t}\cdot p'_{t}}\bigg)\tilde{V}'_{4}(k_{t},p'_{t})\phi_{0P'}(k_{t})+\frac{1}{2}\int\frac{d^{3}k_{t}}{(2\pi)^{3}}[\tilde{V}'_{1}(k_{t},p'_{t})+2(-|p_{t}|\sqrt{\omega^{2}-1}\cos\theta-W_{P}\omega)\tilde{V}'_{2}(k_{t},p'_{t})]\nonumber\\
&&~\times\phi_{0P'}(k_{t})\Bigg]|_{p'_{t}=\sqrt{|p_{t}|^{2}+|p_{t}|^{2}(\omega^{2}-1)\cos^{2}\theta+W^{2}_{P}(\omega^{2}-1)+2\omega W_{P}|p_{t}|\sqrt{\omega^{2}-1}\cos\theta}}.\label{63}
\end{eqnarray}

The three dimensional integrations in Eqs.~(\ref{48}), (\ref{50})-(\ref{61}) can be reduced to one dimensional integrations by using the following identities:
\begin{eqnarray}
&&\int\frac{d^{3}q_{t}}{(2\pi)^{3}}\frac{\rho(q^{2}_{t})}{[(p_{t}-q_{t})^{2}+\mu^{2}]^{2}}=\int\frac{q^{2}_{t}dq_{t}}{4\pi^{2}}\frac{2\rho(q^{2}_{t})}{(p^{2}_{t}+q^{2}_{t}+\mu^{2})^{2}-4p^{2}_{t}q^{2}_{t}},\label{64}\\
&&\int\frac{d^{3}q_{t}}{(2\pi)^{3}}\frac{\rho(q^{2}_{t})}{(p_{t}-q_{t})+\delta^{2}}=\int\frac{q^{2}_{t}dq_{t}}{4\pi^{2}}\frac{\rho(q^{2}_{t})}{2|p_{t}||q_{t}|}ln\frac{(|p_{t}|+|q_{t}|)^{2}+\delta^{2}}{(|p_{t}|-|q_{t}|)^{2}+\delta^{2}},\label{65}
\end{eqnarray}
\noindent where $\rho(q^{2}_{t})$ is some arbitrary function of $q^{2}_{t}$.

\begin{figure}[ht!]
\includegraphics[scale=0.55]{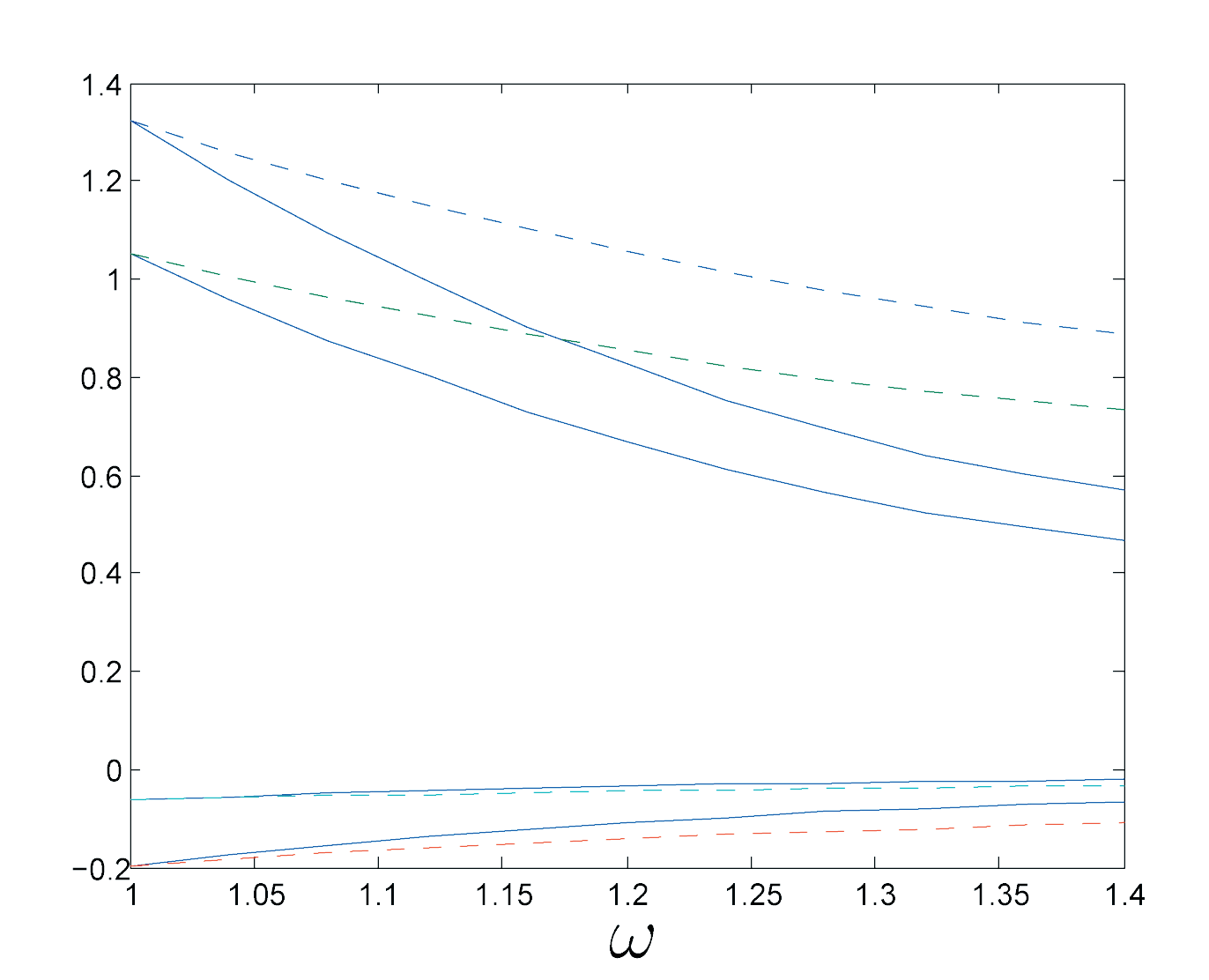}
\includegraphics[scale=0.55]{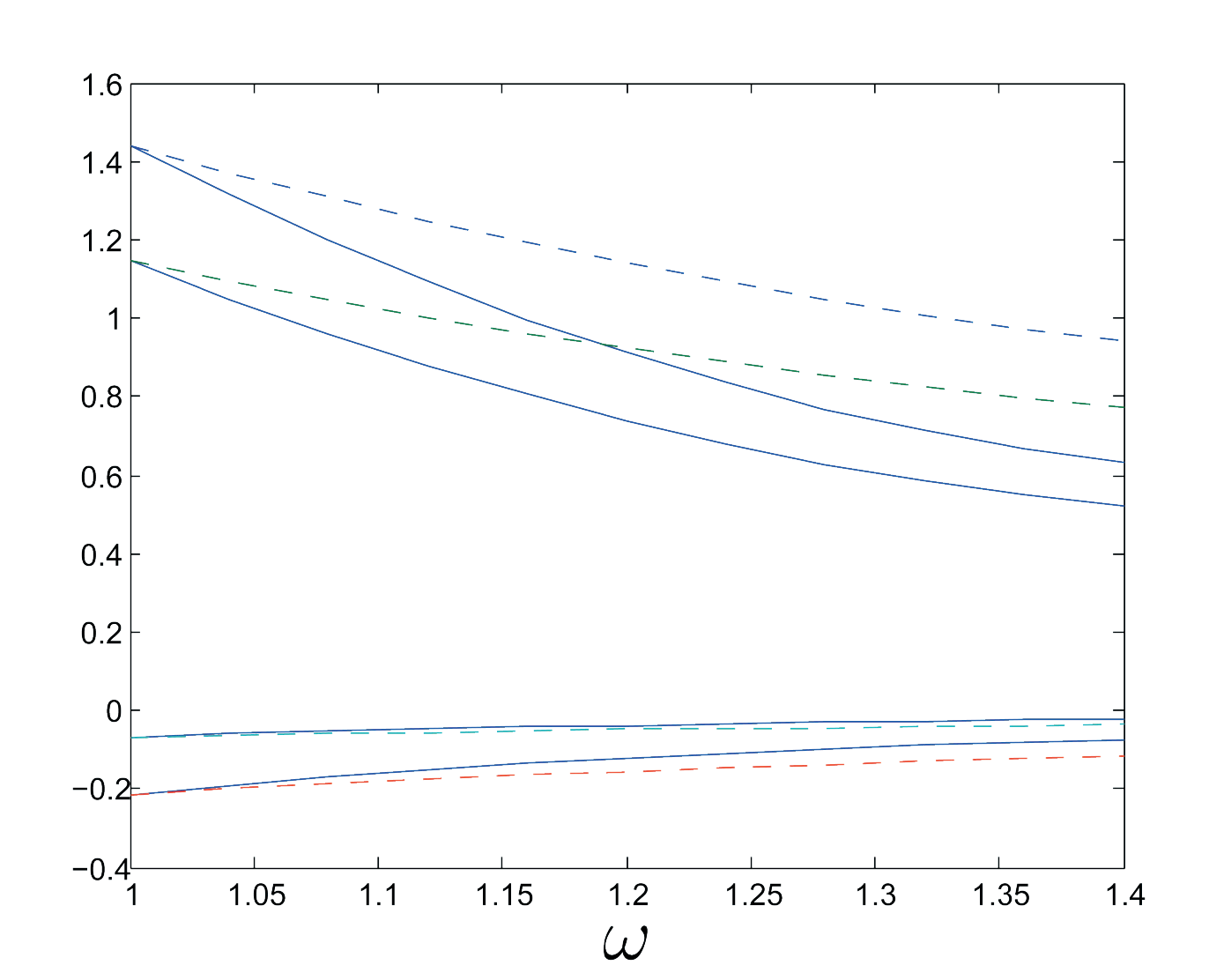}
\caption{The numerical results for $F_i~(i=1,2,3)$ and $G_1$ for $\kappa=0.02GeV^3$~(solid line) and $\kappa=0.08GeV^3$~(dashed line) to the second order in $\frac{1}{m_Q}$ expansion with $m_D=650Mev$~(left one) and $m_D=800MeV$~(right one). From top to bottom we have $F_1$,~$G_1$,~$F_3$ and $F_2$, respectively.}
\label{fig:14}
\end{figure}

\begin{figure}[ht!]
\includegraphics[scale=0.55]{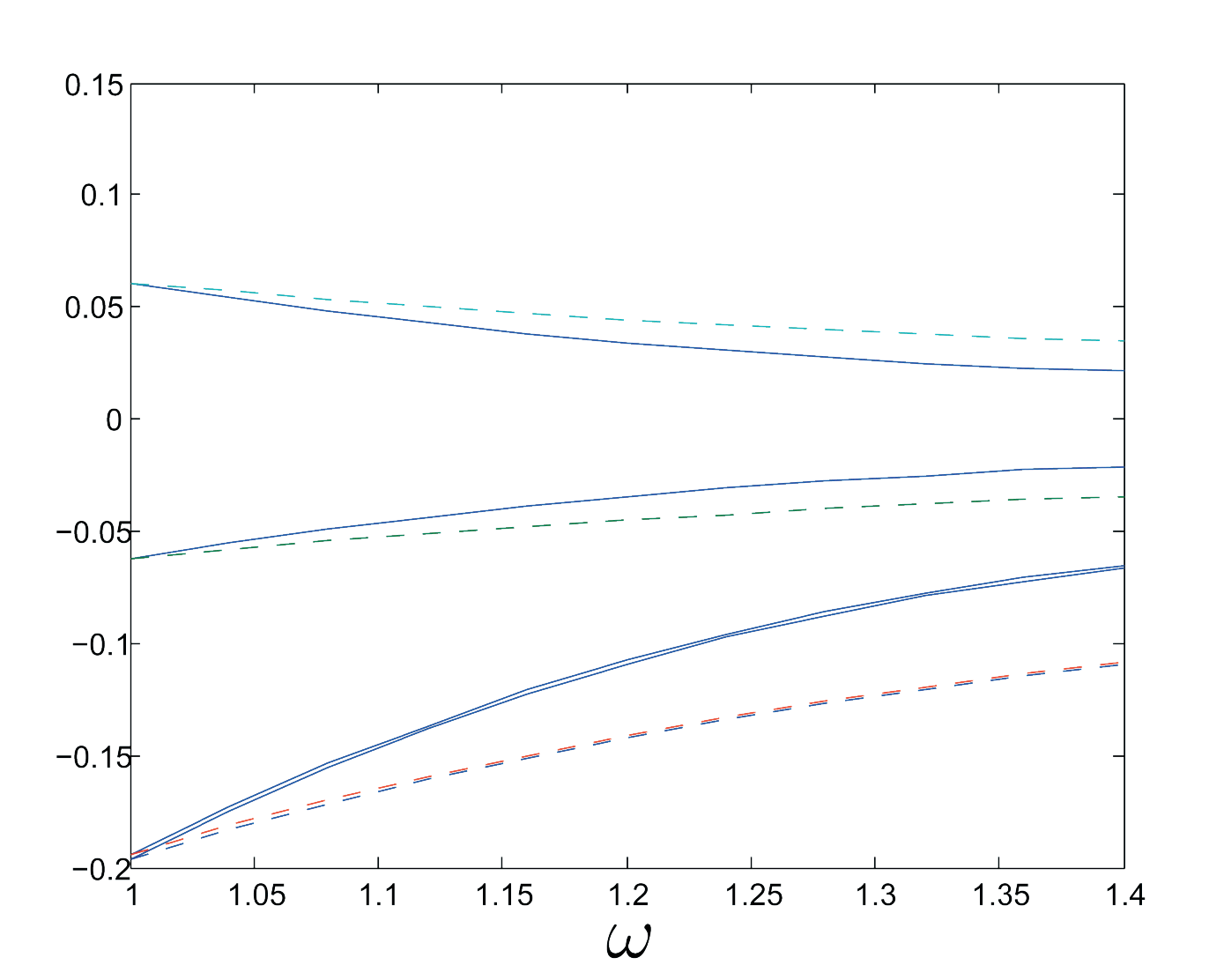}
\includegraphics[scale=0.55]{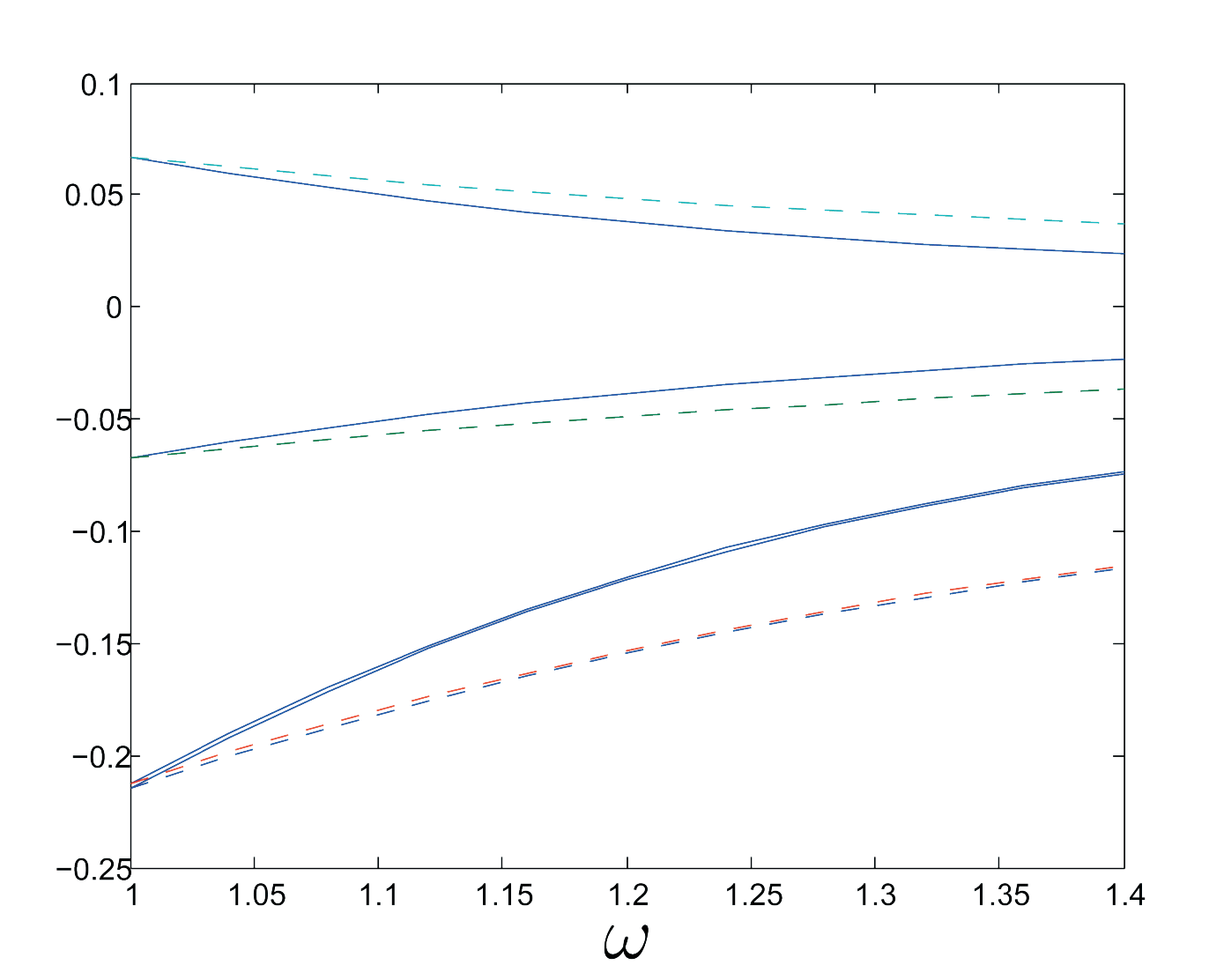}
\caption{The numerical results for $F_2$,~$F_3$,~$G_2$ and $G_3$ for $\kappa=0.02GeV^3$~(solid line) and $\kappa=0.08GeV^3$~(dashed line) to the second order in $\frac{1}{m_Q}$ expansion with $m_D=650Mev$~(left one) and $m_D=800MeV$~(right one). From top to bottom we have $G_3$,~$F_3$,~$F_2$ and $G_2$, We can barely figure out $F_2$ from $G_2$ because they are almost the same.}
\label{fig:15}
\end{figure}

In our model we have several parameters, $\alpha_{seff}$, $\kappa$, $Q^{2}_{0}$, $m_{D}$, $E_{0}$, $E_{1}$ and $E_{2}$. The relationship between $\alpha_{seff}$ and $\kappa$ is independent of the heavy quark mass in the diquark model, hence $\alpha_{seff,\Lambda_{b}}=\alpha_{seff,\Lambda_{c}}$ for the same $\kappa$. As discussed in our previous work, we let $\kappa$ vary between $0.02GeV^{3}$ and $0.08GeV^{3}$. The parameter $Q^{2}_{0}$ can be chosen as $3.2GeV^{2}$ from the data for the electromagnetic form factor of the proton\cite{Ansel1987}. From the BS equation solutions in the meson case, it has been found that the values $m_{b}=5.02GeV$ and $m_{c}=1.58GeV$ give theoretical results which are in good agreement with experiment\cite{Jin1992}. With Eq.~(\ref{14}), It is easy to see that
\begin{eqnarray}
m_{D}+E_{0}+\frac{1}{m_{Q}}E_{1}+\frac{1}{m^{2}_{Q}}E_{2}=0.62GeV,\label{66}
\end{eqnarray}
\noindent if we use $m_{\Lambda_{b}}=5.64GeV$. In order to guarantee that the binding energies $E_{0}$, $E_{1}$ and $E_{2}$ are negative, we assume the minimum value of $m_{D}$ is $650MeV$ and the maximum value of $m_{D}$ is chosen as $800MeV$. One notes that $E_{1}\sim\Lambda_{QCD}E_{0}$, $E_{2}\sim\Lambda_{QCD}E_{1}$, and hence we further assume that $E_{1}=\delta E_{0}$, $E_{2}=\delta E_{1}=\delta^{2}E_{0}$. The value of $m_{D}+E_{0}$ is dependent on $\delta~(0.1-0.4)$. Now with the BS equations to the $1/m^{2}_{Q}$ order in the $1/m_{Q}$ expansion we had in our previous work, we give the numerical results of all the six form factors in Figs.~\ref{fig:14} and \ref{fig:15}.

\section*{IV. APPLICATIONS TO $\Lambda_{b}\rightarrow\Lambda_{c}l\bar{\nu}$ AND $\Lambda_{b}\rightarrow\Lambda_{c}P(V)$  }
\textbf{A. Semileptonic decay $\Lambda_{b}\rightarrow\Lambda_{c}l\bar{\nu}$}\\

When the mass of the lepton can be neglected, the semileptonic decay form factors of baryons $1/2^{+}\rightarrow1/2^{+}$ can be expressed as following:
\begin{eqnarray}
\langle\Lambda_{2}(P_{2})|J^{V+A}_{\mu}|\Lambda_{1}(P_{1})\rangle=\bar{u}(P_{2})\Big[\gamma_{\mu}(F^{V}_{1}+F^{A}_{1}\gamma_{5})+i\sigma_{\mu\nu}q^{\nu}(F^{V}_{2}+F^{A}_{2}\gamma_{5})\Big]u(P_{1}),\label{67}\nonumber\\
\end{eqnarray}
\noindent where $J^{V}_{\mu}$ and $J^{A}_{\mu}$ represent vector current and axial-vector current, respectively, $u(P_{1,2})$ is the Dirac spinor of $\Lambda_{1,2}$, $q_{\mu}=(P_{1}-P_{2})_{\mu}$ is the four momentum transfer, and form factors $F^{V,A}_{i}(i=1,2)$ are functions of $q^{2}$. The terms containing $q^{\mu}$ do not contribute to the semileptonic decay when $m_{lepton}=0$.

Comparing Eq.(\ref{67}) with Eq.(\ref{1}) and noting that $\Lambda_{1}$ is $\Lambda_{b}$ while $\Lambda_{2}$ is $\Lambda_{c}$, we can get the relationships between $F^{V,A}_{1,2}$ and $F_{i}$, $G_{i} (i=1,2,3)$:
\begin{eqnarray}
&&F^{V}_{1}-(m_{\Lambda_{b}}+m_{\Lambda_{c}})F^{V}_{2}=F_{1},\label{68}\\
&&2m_{\Lambda_{c}}F^{V}_{2}=\frac{m_{\Lambda_{c}}}{m_{\Lambda_{b}}}F_{2}+F_{3},\label{69}\\
&&F^{A}_{1}+(m_{\Lambda_{b}}-m_{\Lambda_{c}})F^{A}_{2}=-G_{1},\label{70}\\
&&2m_{\Lambda_{c}}F^{A}_{2}=-\bigg(\frac{m_{\Lambda_{c}}}{m_{\Lambda_{b}}}G_{2}+G_{3}\bigg).\label{71}
\end{eqnarray}

We have to add QCD corrections into the differential decay width because they are comparable with $\frac{1}{m_{Q}}$ correction. Therefore, the form factors are changed to the following\cite{Korner1992}:
\begin{eqnarray}
F^{QCD}_{1}(\omega)=F_{1}(\omega)-iF_{(0P',0P)}\frac{\alpha_{s}}{\pi}v_{1},\label{72}\\
F^{QCD}_{2}(\omega)=F_{2}(\omega)+iF_{(0P',0P)}\frac{\alpha_{s}}{\pi}v_{2},\label{73}\\
F^{QCD}_{3}(\omega)=F_{3}(\omega)+iF_{(0P',0P)}\frac{\alpha_{s}}{\pi}v_{3},\label{74}\\
G^{QCD}_{1}(\omega)=G_{1}(\omega)-iF_{(0P',0P)}\frac{\alpha_{s}}{\pi}a_{1},\label{75}\\
G^{QCD}_{2}(\omega)=G_{2}(\omega)+iF_{(0P',0P)}\frac{\alpha_{s}}{\pi}a_{2},\label{76}\\
G^{QCD}_{3}(\omega)=G_{3}(\omega)+iF_{(0P',0P)}\frac{\alpha_{s}}{\pi}a_{3},\label{77}
\end{eqnarray}
\noindent where $v_{i}$ and $a_{i}(i=1,2,3)$ are QCD corrections\cite{Neubert1992}.

Now with Eqs.~(\ref{68})-(\ref{77}), $F^{V,A}_{1,2}$ with QCD corrections can be expressed as
\begin{eqnarray}
&&F^{V}_{1}(\omega)=F^{QCD}_{1}(\omega)+\bigg(\frac{1}{2}+\frac{m_{\Lambda_{c}}}{2m_{\Lambda_{b}}}\bigg)F^{QCD}_{2}(\omega)+\bigg(\frac{m_{\Lambda_{b}}}{2m_{\Lambda_{c}}}+\frac{1}{2}\bigg)F^{QCD}_{3}(\omega),\label{78}\\
&&F^{V}_{2}(\omega)=\frac{1}{2m_{\Lambda_{b}}}F^{QCD}_{2}(\omega)+\frac{1}{2m_{\Lambda_{c}}}F^{QCD}_{3}(\omega),\label{79}\\
&&F^{A}_{1}(\omega)=-G^{QCD}_{1}(\omega)+\bigg(\frac{1}{2}-\frac{m_{\Lambda_{c}}}{2m_{\Lambda_{b}}}\bigg)G^{QCD}_{2}(\omega)+\bigg(\frac{m_{\Lambda_{b}}}{2m_{\Lambda_{c}}}-\frac{1}{2}\bigg)G^{QCD}_{3}(\omega),\label{80}\\
&&F^{A}_{2}(\omega)=-\frac{1}{2m_{\Lambda_{b}}}G^{QCD}_{2}(\omega)-\frac{1}{2m_{\Lambda_{c}}}G^{QCD}_{3}(\omega).\label{81}
\end{eqnarray}

$v_i$ and $a_i~(i=1,2,3)$ can be expressed respectively as\cite{Neubert1992}:\\
\begin{eqnarray}
v_{1}&=&\frac{1}{3}\Big(K_{V}+2D^{V}_{1}+2D^{V}_{2}+2(\omega-1)D^{V}_{3}\Big),\label{82}\\
v_{2}&=&\frac{2}{3}D^{V}_{2},\label{83}\\
v_{3}&=&\frac{2}{3}D^{V}_{1},\label{84}\\
a_{1}&=&\frac{1}{3}\Big(K_{A}-2(\omega-1)D^{A}_{3}\Big),\label{85}\\
a_{2}&=&\frac{2}{3}\Big(D^{A}_{2}-2D^{A}_{3}\Big),\label{86}\\
a_{3}&=&-\frac{2}{3}\Big(D^{A}_{1}-2D^{A}_{3}\Big),\label{87}
\end{eqnarray}

\noindent where\\
\begin{eqnarray}
K_{V}&=&-\Big(2\big[\omega r(\omega)-1\big]ln\frac{m_{\Lambda_{b}}m_{\Lambda_{c}}}{\lambda^{2}}+C_{V}(Z,\omega)\Big),\label{88}\\
K_{A}&=&-\Big(2\big[\omega r(\omega)-1\big]ln\frac{m_{\Lambda_{b}}m_{\Lambda_{c}}}{\lambda^{2}}+C_{A}(Z,\omega)\Big),\label{89}\\
D^{V}_{1}&=&\Big(1-\frac{1}{2}Z\Big)r(\omega)-\Big(1+\frac{1}{2}Z\Big)\chi(\omega)-\frac{Z}{1-Z}\phi(\omega),\label{90}\\
D^{V}_{2}&=&\Big(1-\frac{1}{2Z}\Big)r(\omega)-\Big(1+\frac{1}{2Z}\Big)\chi(\omega)-\frac{1}{Z(Z-1)}\phi(\omega),\label{91}\\
D^{V}_{3}&=&-\frac{Z}{(1-Z)^{2}}\phi(\omega),\label{92}\\
D^{A}_{1}&=&\Big(1+\frac{1}{2}Z\Big)r(\omega)-\Big(1-\frac{1}{2}Z\Big)\chi(\omega)+\frac{Z}{1-Z}\phi(\omega),\label{93}\\
D^{A}_{2}&=&\Big(1+\frac{1}{2Z}\Big)r(\omega)-\Big(1-\frac{1}{2Z}\Big)\chi(\omega)+\frac{1}{Z(Z-1)}\phi(\omega),\label{94}\\
D^{A}_{3}&=&\frac{Z}{(1-Z)^{2}}\phi(\omega),\label{95}
\end{eqnarray}
\noindent with the following definitions:\\
\begin{eqnarray}
r(\omega)&=&\frac{1}{\sqrt{\omega^{2}-1}}ln(\omega+\sqrt{\omega^{2}-1}),\label{96}\\
Z&=&\frac{m_{c}}{m_{b}},\label{97}\\
C_{V}(Z,\omega)&=&5+G(\omega)+2H(\omega)-2(2\omega-1)r(\omega)-\frac{3}{2}\Big(\frac{1+Z^{2}}{Z}r(\omega)-\frac{1-Z^{2}}{Z}\chi(\omega)\Big)+\phi(\omega),\label{98}\\
C_{A}(Z,\omega)&=&5+G(\omega)+2H(\omega)-2(2\omega-1)r(\omega)-\frac{1}{2}\Big(\frac{1+Z^{2}}{Z}r(\omega)-\frac{1-Z^{2}}{Z}\chi(\omega)\Big)-\phi(\omega),\label{99}\\
\chi(\omega)&=&\frac{1}{1-2Z\omega+Z^{2}}\Big(2ZlnZ+(1-Z^{2})r(\omega)\Big),\label{100}\\
\phi(\omega)&=&\frac{(1-Z)^{2}}{1-2Z\omega+Z^{2}}\Big(G(\omega)+1-\omega r(\omega)\Big),\label{101}\\
G(\omega)&=&\frac{1}{1-2Z\omega+Z^{2}}\Big((Z^{2}-1)lnZ-2Z(\omega^{2}-1)r(\omega)\Big)-2,\label{102}\\
H(\omega)&=&\omega r(\omega)lnZ+h(\omega)+2-\frac{\omega}{\sqrt{\omega^{2}-1}}\Big[L_{2}(1-Zy_{-})-L_{2}(1-Zy_{+})\Big],\label{103}\\
y_{\pm}&=&\omega\pm\sqrt{\omega^{2}-1},\label{104}\\
L_{2}(x)&=&-\int^{x}_{0}dt\frac{ln(1-t)}{t},\label{105}\\
h(\omega)&=&\frac{\omega}{2\sqrt{\omega^{2}-1}}\Big[L_{2}(1-y^{2}_{-})-L_{2}(1-y^{2}_{+})\Big].\label{106}
\end{eqnarray}

In Ref.~\cite{Korner1992}, K$\ddot{o}$rner and Kr$\ddot{a}$mer regarded the semileptonic decay $\Lambda_1\rightarrow\Lambda_2+l+\nu_l$ as a quasi two-body decay $\Lambda_1\rightarrow\Lambda_2+W_{off-shell}$ followed by the leptonic decay $W_{off-shell}\rightarrow l+\nu_l$. In the zero-lepton-mass approximation only the $J^P=1^+,~1^-$ components of $W_{off-shell}$ participate in the decay~($1^+$ corresponds to axial-vector current while $1^-$ to vector current). They defined helicity amplitudes $H^{V,A}_{\lambda_2~\lambda_W}$ where $\lambda_2$ and $\lambda_W$ are the helicities of the daughter baryon and the $W_{off-shell}$ boson, respectively. They are related to the invariant form factors through\cite{Korner1992}:\\
\begin{eqnarray}
\sqrt{q^{2}}H^{V}_{\frac{1}{2}~0}=\sqrt{Q_{-}}\Big[(M_{1}+M_{2})F^{V}_{1}-q^{2}F^{V}_{2}\Big],\label{107}\\
H^{V}_{\frac{1}{2}~1}=\sqrt{2Q_{-}}\Big[-F^{V}_{1}+(M_{1}+M_{2})F^{V}_{2}\Big],\label{108}\\
\sqrt{q^{2}}H^{A}_{\frac{1}{2}~0}=\sqrt{Q_{+}}\Big[(M_{1}-M_{2})F^{A}_{1}-q^{2}F^{A}_{2}\Big],\label{109}\\
H^{A}_{\frac{1}{2}~1}=\sqrt{2Q_{+}}\Big[-F^{V}_{1}-(M_{1}-M_{2})F^{A}_{2}\Big],\label{110}
\end{eqnarray}
\noindent where $Q_{\pm}=(M_1\pm M_2)^2-q^2$ with $M_1$ and $M_2$ being the masses of $\Lambda_1$ and $\Lambda_2$, respectively. The remaining helicity amplitudes can be obtained with the help of the parity relations:
\begin{eqnarray}
H^{V}_{-\lambda_{2}~-\lambda_{W}}=+H^{V}_{\lambda_{2}~\lambda_{W}},\label{111}\\
H^{A}_{-\lambda_{2}~-\lambda_{W}}=-H^{A}_{\lambda_{2}~\lambda_{W}}.\label{112}
\end{eqnarray}

The differential decay width of the semileptonic decay $\Lambda_b\rightarrow\Lambda_c l\bar{\nu}$ is
\begin{eqnarray}
\frac{d\Gamma}{Adq^{2}}&=&2\cdot\frac{q^{2}p}{48m^{2}_{\Lambda_{b}}}\bigg(|H^{V}_{\frac{1}{2}~1}+H^{A}_{\frac{1}{2}~1}|^{2}+|H^{V}_{-\frac{1}{2}~-1}+H^{A}_{-\frac{1}{2}~-1}|^{2}\nonumber\\
&&~~~~~~~~~+|H^{V}_{\frac{1}{2}~0}+H^{A}_{\frac{1}{2}~0}|^{2}+|H^{V}_{-\frac{1}{2}~0}+H^{A}_{-\frac{1}{2}~0}|^{2}\bigg),\label{113}
\end{eqnarray}
\noindent where $p$ is the norm of the three dimensional momentum of daughter $\Lambda_c$. $p$ and $A$ can be expressed as:
\begin{eqnarray}
&&p=\frac{\sqrt{m^{4}_{\Lambda_{b}}+m^{4}_{\Lambda_{c}}+q^{4}-2m^{2}_{\Lambda_{b}}m^{2}_{\Lambda_{c}}-2m^{2}_{\Lambda_{b}}q^{2}-2m^{2}_{\Lambda_{c}}q^{2}}}{2m_{\Lambda_{b}}}\nonumber\\
&&~~~=m_{\Lambda_{c}}\sqrt{\omega^{2}-1},\label{114}\\
&&A=\frac{G^{2}_{F}}{(2\pi)^{3}}|V_{cb}|^{2}B(\Lambda_{c}\rightarrow ab),\label{115}
\end{eqnarray}
\noindent where $V_{cb}$ is the Cabibbo-Kobayashi-Maskawa matrix element and $B(\Lambda_c\rightarrow ab)$ is two-body decay branching ratio of $\Lambda_c\rightarrow ab$.

With the help of $dq^{2}=2m_{\Lambda_{b}}m_{\Lambda_{c}}d\omega$, one can get the differential decay width in terms of  $\omega$:
\begin{eqnarray}
\frac{d\Gamma}{Ad\omega}&=&\frac{m^{2}_{\Lambda_{c}}\sqrt{\omega^{2}-1}q^{2}}{12m_{\Lambda_{b}}}\bigg(|H^{V}_{\frac{1}{2}~1}+H^{A}_{\frac{1}{2}~1}|^{2}+|H^{V}_{-\frac{1}{2}~-1}+H^{A}_{-\frac{1}{2}~-1}|^{2}\nonumber\\
&&~~~~~~~~~+|H^{V}_{\frac{1}{2}~0}+H^{A}_{\frac{1}{2}~0}|^{2}+|H^{V}_{-\frac{1}{2}~0}+H^{A}_{-\frac{1}{2}~0}|^{2}\bigg)\nonumber\\
&=&\frac{m^{2}_{\Lambda_{c}}\sqrt{\omega^{2}-1}q^{2}}{6m_{\Lambda_{b}}}\bigg(|H^{V}_{\frac{1}{2}~1}|^{2}+|H^{A}_{\frac{1}{2}~1}|^{2}+|H^{V}_{\frac{1}{2}~0}|^{2}+|H^{A}_{\frac{1}{2}~0}|^{2}\bigg).\label{116}
\end{eqnarray}

With Eqs.~(\ref{107})-(\ref{110}), one can immediately get
\begin{eqnarray}
&&\sum|H|^{2}\nonumber\\
&=&\bigg(|H^{V}_{\frac{1}{2}~1}|^{2}+|H^{A}_{\frac{1}{2}~1}|^{2}+|H^{V}_{\frac{1}{2}~0}|^{2}+|H^{A}_{\frac{1}{2}~0}|^{2}\bigg)\nonumber\\
&=&2m_{\Lambda_{b}}m_{\Lambda_{c}}\Bigg\{\bigg(\frac{2\omega+2\omega\eta^{2}-4\eta}{1+\eta^{2}-2\omega\eta}+4\omega\bigg)(-F^{2}_{(0P',0P)})\nonumber\\
&&~~~~+\frac{1}{m_{b}}\Bigg[-2\bigg(\frac{2\omega+2\omega\eta^{2}-4\eta}{1+\eta^{2}-2\omega\eta}+4\omega\bigg)(f_{1(0P',1P^{-})}+\eta^{-1}f_{2(0P',1P^{-})})F_{(0P',0P)}\nonumber\\
&&~~~~~~~~~~+12\eta^{-1}(\omega-\eta)f_{2(0P',1P^{-})}F_{(0P',0P)}\Bigg]\nonumber\\
&&~~~~+\frac{1}{m_{c}}\Bigg[-2\bigg(\frac{2\omega+2\omega\eta^{2}-4\eta}{1+\eta^{2}-2\omega\eta}+4\omega\bigg)(f_{4(1P'^{-},0P)}+\eta f_{3(1P'^{-},0P)})F_{(0P',0P)}\nonumber\\
&&~~~~~~~~~~+12(\omega\eta-1)f_{3(1P'^{-},0P)}F_{(0P',0P)}\Bigg]\nonumber\\
&&~~~~+\frac{1}{m^{2}_{b}}\Bigg[-\bigg(\frac{2\omega+2\omega\eta^{2}-4\eta}{1+\eta^{2}-2\omega\eta}+4\omega\bigg)\Big(f^{2}_{1(0P',1P^{-})}+\eta^{-2}f^{2}_{2(0P',1P^{-})}\nonumber\\
&&~~~~~~~~~~+2\eta^{-1}f_{1(0P',1P^{-})}f_{2(0P',1P^{-})}+2(F_{(0P',2P^{+})}+f_{1(0P',2P^{-})}+\eta^{-1}f_{2(0P',1P^{-})})\nonumber\\
&&~~~~~~~~~~\times F_{(0P',0P)}\Big)-\eta^{-2}\Big(2\omega(3+\eta^{2}-2\omega\eta+2\eta^{2})-8\eta\Big)f^{2}_{2(0P',1P^{-})}\nonumber\\
&&~~~~~~~~~~+12\eta(\omega-\eta)(f_{1(0P',1P^{-})}+\eta^{-1}f_{2(0P',1P^{-})})f_{2(0P',1P^{-})}\Bigg]\nonumber\\
&&~~~~+\frac{1}{m^{2}_{c}}\Bigg[-\bigg(\frac{2\omega+2\omega\eta^{2}-4\eta}{1+\eta^{2}-2\omega\eta}+4\omega\bigg)\Big(f^{2}_{4(1P'^{-},0P)}+\eta^{2}f_{3(1P'^{-},0P)}\nonumber\\
&&~~~~~~~~~~+2\eta f_{4(1P'^{-},0P)}f_{3(1P'^{-},0P)}+2(F_{(2P'^{+},0P)}+f_{4(2P'^{-},0P)}+\eta f_{3(2P'^{-},0P)})\nonumber\\
&&~~~~~~~~~~\times F_{(0P',0P)}\Big)-\Big(2\omega(3+\eta^{2}-2\omega\eta+2\eta^{2})-8\eta\Big)f^{2}_{3(1P'^{-},0P)}\nonumber\\
&&~~~~~~~~~~+12(\omega-\eta)(f_{4(1P'^{-},0P)}+\eta f_{3(1P'^{-},0P)})f_{3(1P'^{-},0P)}\Bigg]\nonumber\\
&&~~~~+\frac{1}{m_{b}m_{c}}\Bigg[-2\eta\bigg(\frac{2\omega+2\omega\eta^{2}-4\eta}{1+\eta^{2}-2\omega\eta}+4\omega\bigg)f_{7(1P'^{-},1P^{-})}F_{(0P',0P)}\nonumber\\
&&~~~~~~~~~~-2\bigg(\frac{4\omega\eta-2\eta^{2}-2}{1+\eta^{2}-2\eta\omega}-4\bigg)(1+\eta^{-1})f_{7(1P'^{-},1P^{-})}F_{(0P',0P)}\nonumber\\
&&~~~~~~~~~~-2\bigg(\frac{2\omega+2\omega\eta^{2}-4\eta}{1+\eta^{2}-2\omega\eta}+4\omega\bigg)\big(f_{1(0P',1P^{-})}f_{4(1P'^{-},0P)}+\eta f_{1(0P',1P^{-})}f_{3(1P'^{-},0P)}\nonumber\\
&&~~~~~~~~~~+\eta^{-1}f_{2(0P',1P^{-})}f_{4(1P'^{-},0P)}+f_{2(0P',1P^{-})}f_{3(1P'^{-},0P)}\big)\nonumber\\ &&~~~~~~~~~~+2\eta^{-1}\Big(2(3+\eta^{2}-2\omega\eta+2\eta^{2})-8\omega\eta\Big)f_{3(1P'^{-},0P)}f_{2(0P',1P^{-})}\nonumber\\
&&~~~~~~~~~~+12(\omega\eta-1)(f_{1(0P',1P^{-})}+\eta^{-1}f_{2(0P',1P^{-})})f_{3(1P'^{-},0P)}+12\eta^{-1}(\omega-\eta)\nonumber\\
&&~~~~~~~~~~\times(f_{4(1P'^{-},0P)}+\eta f_{3(1P'^{-},0P)})f_{2(0P',1P^{-})}\Bigg]\nonumber
\end{eqnarray}
\begin{eqnarray}
&&~~~~+\frac{\alpha_{s}}{\pi}v_{1}(\omega-1)F_{(0P',0P)}\Bigg[-2\bigg(\frac{(1+\eta)^{2}}{1+\eta^{2}-2\omega\eta}+2\bigg)\Big(F_{(0P',0P)}+\frac{1}{m_{b}}(f_{1(0P',1P^{-})}\nonumber\\
&&~~~~~~~~~~+\eta^{-1}f_{2(0P',1P^{-})})+\frac{1}{m_{c}}(f_{4(1P'^{-},0P)}+\eta f_{3(1P'^{-},0P)})\Big)+6(1+\eta)\big(\eta^{-1}\frac{1}{m_{b}}f_{2(0P',1P^{-})}\nonumber\\
&&~~~~~~~~~~+\frac{1}{m_{c}}f_{3(1P'^{-},0P)}\big)\Bigg]\nonumber\\
&&~~~~+\frac{\alpha_{s}}{\pi}v_{2}(\omega-1)F_{(0P',0P)}\Bigg[2\bigg(\frac{(1+\eta)^{2}}{1+\eta^{2}-2\omega\eta}+2\bigg)\Big(F_{(0P',0P)}+\frac{1}{m_{b}}(f_{1(0P',1P^{-})}\nonumber\\
&&~~~~~~~~~~+\eta^{-1}f_{2(0P',1P^{-})})+\frac{1}{m_{c}}(f_{4(1P'^{-},0P)}+\eta f_{3(1P'^{-},0P)})\Big)\Big(\frac{1+\eta}{2}\Big)-6(1+\eta)\nonumber\\ &&~~~~~~~~~~\times\Big[\eta^{-1}\frac{1}{m_{b}}\big(\frac{1+\eta}{2}\big)f_{2(0P',1P^{-})}+\frac{1}{m_{c}}\big(\frac{1+\eta}{2}\big)f_{3(1P'^{-},0P)}+\frac{1}{2}\Big(F_{(0P',0P)}\nonumber\\ &&~~~~~~~~~~+\frac{1}{m_{b}}(f_{1(0P',1P^{-})}+\eta^{-1}f_{2(0P',1P^{-})})+\frac{1}{m_{c}}(f_{4(1P'^{-},0P)}+\eta f_{3(1P'^{-},0P)})\Big)\Big]\nonumber\\
&&~~~~~~~~~~+\big(1+\eta^{2}-2\omega\eta+2(1+\eta)^{2}\big)\Big(\eta^{-1}\frac{1}{m_{b}}f_{2(0P',1P^{-})}+\frac{1}{m_{c}}f_{3(1P'^{-},0P)}\Big)\Bigg]\nonumber\\
&&~~~~+\frac{\alpha_{s}}{\pi}v_{3}(\omega-1)F_{(0P',0P)}\Bigg[2\bigg(\frac{(1+\eta)^{2}}{1+\eta^{2}-2\omega\eta}+2\bigg)\Big(F_{(0P',0P)}+\frac{1}{m_{b}}(f_{1(0P',1P^{-})}\nonumber\\
&&~~~~~~~~~~+\eta^{-1}f_{2(0P',1P^{-})})+\frac{1}{m_{c}}(f_{4(1P'^{-},0P)}+\eta f_{3(1P'^{-},0P)})\Big)\Big(\frac{1+\eta}{2\eta}\Big)-6(1+\eta)\nonumber\\ &&~~~~~~~~~~\times\Big[\eta^{-1}\frac{1}{m_{b}}\big(\frac{1+\eta}{2\eta}\big)f_{2(0P',1P^{-})}+\frac{1}{m_{c}}\big(\frac{1+\eta}{2\eta}\big)f_{3(1P'^{-},0P)}+\frac{\eta^{-1}}{2}\Big(F_{(0P',0P)}\nonumber\\ &&~~~~~~~~~~+\frac{1}{m_{b}}(f_{1(0P',1P^{-})}+\eta^{-1}f_{2(0P',1P^{-})})+\frac{1}{m_{c}}(f_{4(1P'^{-},0P)}+\eta f_{3(1P'^{-},0P)})\Big)\Big]\nonumber\\
&&~~~~~~~~~~+\big(1+\eta^{2}-2\omega\eta+2(1+\eta)^{2}\big)\Big(\eta^{-1}\frac{1}{m_{b}}f_{2(0P',1P^{-})}+\frac{1}{m_{c}}f_{3(1P'^{-},0P)}\Big)\Bigg]\nonumber\\
&&~~~~+\frac{\alpha_{s}}{\pi}a_{1}(\omega+1)F_{(0P',0P)}\Bigg[-2\bigg(\frac{(1-\eta)^{2}}{1+\eta^{2}-2\omega\eta}+2\bigg)\Big(F_{(0P',0P)}+\frac{1}{m_{b}}(f_{1(0P',1P^{-})}\nonumber\\
&&~~~~~~~~~~+\eta^{-1}f_{2(0P',1P^{-})})+\frac{1}{m_{c}}(f_{4(1P'^{-},0P)}+\eta f_{3(1P'^{-},0P)})\Big)+6(1-\eta)\big(\eta^{-1}\frac{1}{m_{b}}f_{2(0P',1P^{-})}\nonumber\\
&&~~~~~~~~~~-\frac{1}{m_{c}}f_{3(1P'^{-},0P)}\big)\Bigg]\nonumber\\
&&~~~~+\frac{\alpha_{s}}{\pi}a_{2}(\omega+1)F_{(0P',0P)}\Bigg[2\bigg(\frac{(1-\eta)^{2}}{1+\eta^{2}-2\omega\eta}+2\bigg)\Big(F_{(0P',0P)}+\frac{1}{m_{b}}(f_{1(0P',1P^{-})}\nonumber\\
&&~~~~~~~~~~+\eta^{-1}f_{2(0P',1P^{-})})+\frac{1}{m_{c}}(f_{4(1P'^{-},0P)}+\eta f_{3(1P'^{-},0P)})\Big)\Big(\frac{\eta-1}{2}\Big)-6(\eta-1)\nonumber\\ &&~~~~~~~~~~\times\Big[\eta^{-1}\frac{1}{m_{b}}\big(\frac{1-\eta}{2}\big)f_{2(0P',1P^{-})}-\frac{1}{m_{c}}\big(\frac{1-\eta}{2}\big)f_{3(1P'^{-},0P)}+\frac{1}{2}\Big(F_{(0P',0P)}\nonumber\\ &&~~~~~~~~~~+\frac{1}{m_{b}}(f_{1(0P',1P^{-})}+\eta^{-1}f_{2(0P',1P^{-})})+\frac{1}{m_{c}}(f_{4(1P'^{-},0P)}+\eta f_{3(1P'^{-},0P)})\Big)\Big]\nonumber
\end{eqnarray}
\begin{eqnarray}
&&~~~~~~~~~~+\big(1+\eta^{2}-2\omega\eta+2(1-\eta)^{2}\big)\Big(-\eta^{-1}\frac{1}{m_{b}}f_{2(0P',1P^{-})}+\frac{1}{m_{c}}f_{3(1P'^{-},0P)}\Big)\Bigg]\nonumber\\
&&~~~~+\frac{\alpha_{s}}{\pi}a_{3}(\omega+1)F_{(0P',0P)}\Bigg[2\bigg(\frac{(1-\eta)^{2}}{1+\eta^{2}-2\omega\eta}+2\bigg)\Big(F_{(0P',0P)}+\frac{1}{m_{b}}(f_{1(0P',1P^{-})}\nonumber\\
&&~~~~~~~~~~+\eta^{-1}f_{2(0P',1P^{-})})+\frac{1}{m_{c}}(f_{4(1P'^{-},0P)}+\eta f_{3(1P'^{-},0P)})\Big)\Big(\frac{\eta-1}{2\eta}\Big)-6(\eta-1)\nonumber\\ &&~~~~~~~~~~\times\Big[\eta^{-1}\frac{1}{m_{b}}\big(\frac{1-\eta}{2\eta}\big)f_{2(0P',1P^{-})}-\frac{1}{m_{c}}\big(\frac{1-\eta}{2\eta}\big)f_{3(1P'^{-},0P)}+\frac{\eta^{-1}}{2}\Big(F_{(0P',0P)}\nonumber\\ &&~~~~~~~~~~+\frac{1}{m_{b}}(f_{1(0P',1P^{-})}+\eta^{-1}f_{2(0P',1P^{-})})+\frac{1}{m_{c}}(f_{4(1P'^{-},0P)}+\eta f_{3(1P'^{-},0P)})\Big)\Big]\nonumber\\
&&~~~~~~~~~~+\big(1+\eta^{2}-2\omega\eta+2(1-\eta)^{2}\big)\Big(\eta^{-1}\frac{1}{m_{b}}f_{2(0P',1P^{-})}+\frac{1}{m_{c}}f_{3(1P'^{-},0P)}\Big)\Bigg]\Bigg\}\label{284}\nonumber\\
&&~~~~+O\Big(\frac{1}{m^{3}_{Q}}\Big)+O\Big(\alpha^{2}_{s}\Big)+O\Big(\frac{\alpha_{s}}{m^{2}_{Q}}\Big),\label{117}\nonumber\\
\end{eqnarray}
\noindent where $\eta=m_{\Lambda_c}/m_{\Lambda_b}$.

\begin{figure}[!ht]
\includegraphics[scale=0.55]{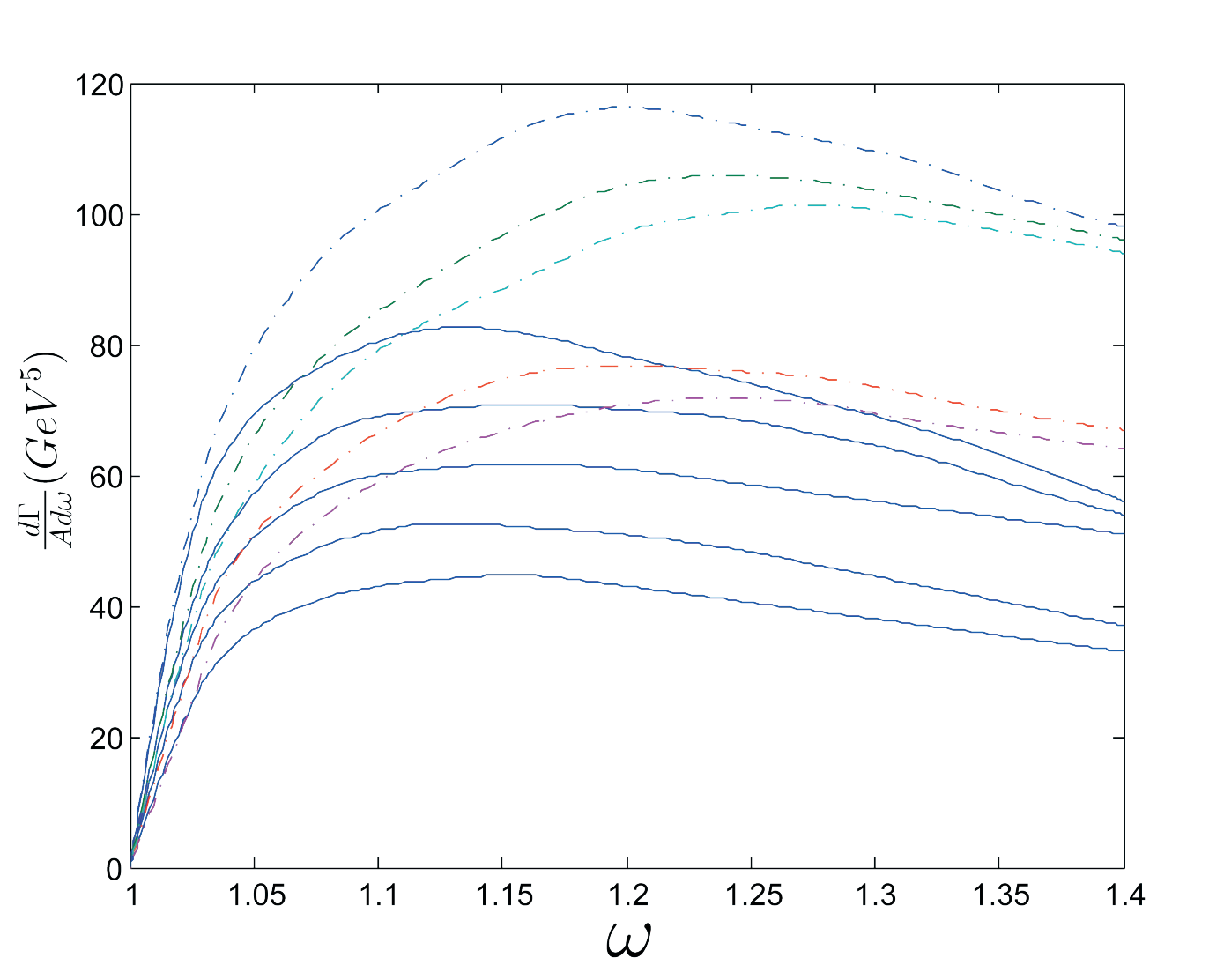}
\includegraphics[scale=0.55]{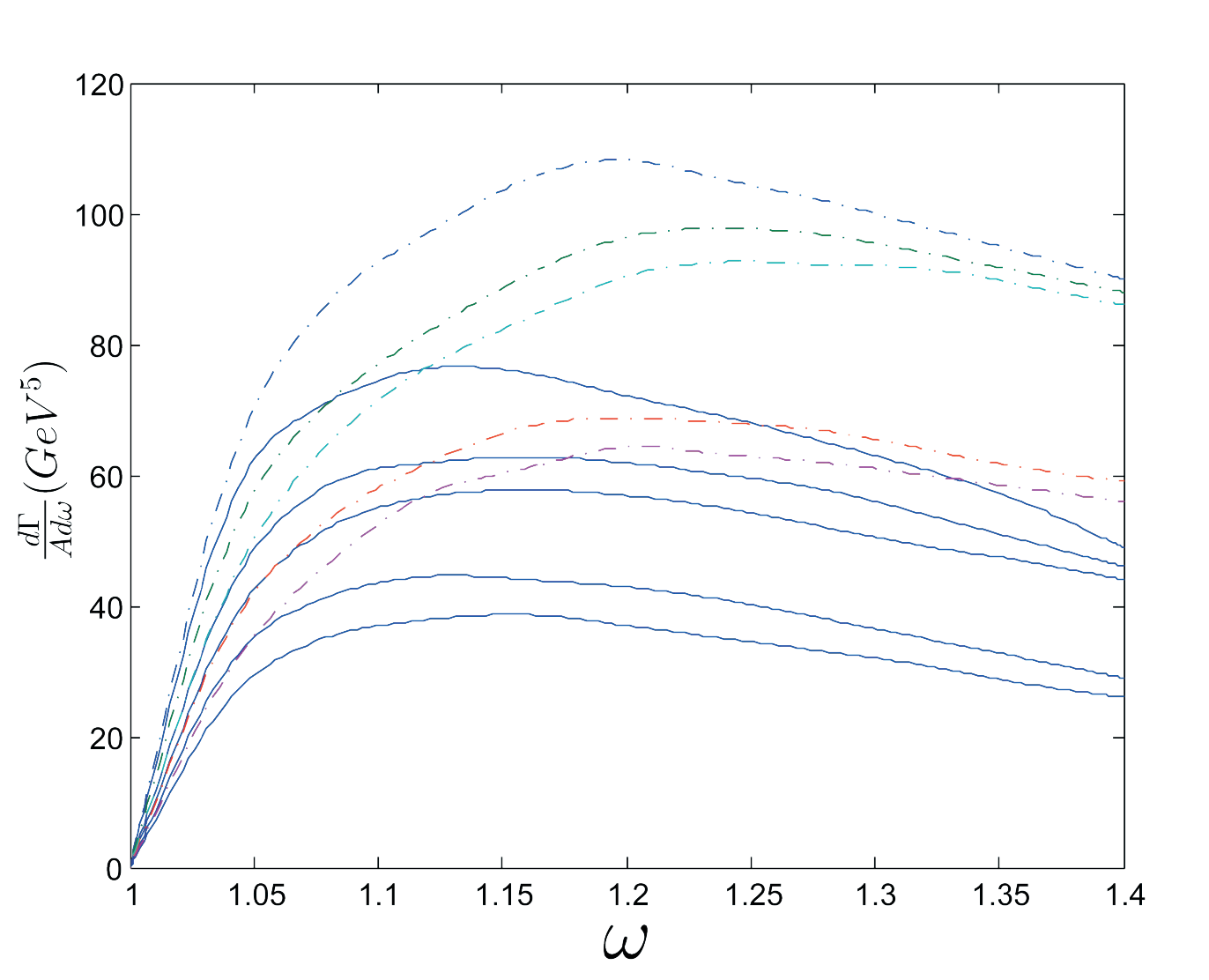}
\caption{The numerical results for $A^{-1}(d\Gamma/d\omega)$ for $\kappa=0.02GeV^3$~(solid lines) and $\kappa=0.08GeV^3$~(dot-dashed lines) with $m_{D}=0.65GeV$~(left one) and $m_D=0.8GeV$~(right one). From top to bottom we have the predictions without $1/m_Q$ and $QCD$ corrections, with $1/m_Q$ corrections, with $1/m_Q+1/m^2_Q$ corrections, with $1/m_Q$ and $QCD$ corrections, and with $1/m_Q+1/m^2_Q$ and $QCD$ corrections, respectively.}
\label{fig:4}
\end{figure}

By using the numerical results of form factors we get in Section III, and substituting Eqs. (\ref{82})-(\ref{106}) and (\ref{117}) into Eq. (\ref{116}), one can get the plots for $\frac{d\Gamma}{Ad\omega}$ shown in Fig. \ref{fig:4} for $m_D=0.65GeV$ and $m_D=0.8GeV$, where we also show explicitly the effects of $1/m_Q$, $1/m^2_Q$ and $QCD$ corrections. For other values of $m_D$ the results change only a little.

We have the total decay width for $\Lambda_b\rightarrow\Lambda_c l\bar{\nu}$ after integrating over $\omega$. The numerical results are shown in Table \ref{tab:1} for $m_D=0.65GeV$ and $m_D=0.8GeV$, and for $\kappa=0.02GeV^3$~($\kappa=0.08GeV^3$). $\Gamma_0$, $\Gamma_{1/m_Q}$, $\Gamma_{1/m_Q+1/m^2_Q}$, $\Gamma_{1/m_Q+QCD}$ and $\Gamma_{1/m_Q+1/m^2_Q+QCD}$ are the decay widths without $1/m_Q$ and $QCD$ corrections, with $1/m_Q$ corrections, with $1/m_Q+1/m^2_Q$ corrections, with $1/m_Q$ and $QCD$ corrections, and with $1/m_Q+1/m^2_Q$ and $QCD$ corrections, respectively.
\begin{table}
\begin{center}
\caption{Predictions for the decay widths for $\Lambda_{b}\rightarrow\Lambda_{c}l\bar{\nu}$ in units $10^{10}s^{-1}B(\Lambda_c\rightarrow ab)$ for $V_{cb}=0.0398/0.042$\cite{PDG14}.}
\begin{tabular}{|c|c c c|}\hline
$m_{D}(GeV)$ & $\Gamma_{0}$ & $\Gamma_{1/m_{Q}}$ & $\Gamma_{1/m_{Q}+1/m^{2}_{Q}}$  \\ \hline 
$0.65$  & $3.74/4.17(5.39/6.01)$ & $3.35/3.72(4.92/5.47)$ & $3.15/3.49(4.65/5.18)$ \\
\hline $0.80$ & $4.34/4.78(5.18/5.76.)$ &
$3.87/4.36(4.76/5.28)$ & $3.66/4.08(4.59/5.09)$  \\ 
\hline $m_{D}(GeV)$ & $\Gamma_{1/m_{Q}+QCD}$ & $\Gamma_{1/m_{Q}+1/m^{2}_{Q}+QCD}$ & \\
\hline $0.65$ & $2.41/2.68(3.58/3.98)$ & $2.15/2.46(3.28/3.65)$ & \\
\hline $0.80$ & $2.79/3.06(3.46/3.84)$ & $2.48/2.88(3.24/3.60)$ & \\
\hline
\end{tabular}\label{tab:1}
\end{center}
\end{table}

From Figure \ref{fig:4}, it is clear to see that $1/m_Q$ corrections, $1/m^2_Q$ corrections and $QCD$ corrections all reduce the differential decay width. From Table \ref{tab:1} it is easy to see that the decay width of $\Lambda_b\rightarrow\Lambda_c l\bar{\nu}$ is $(2.15\sim3.60)\times10^{10}s^{-1}B(\Lambda_c\rightarrow ab)$ after taking $1/m_Q$ corrections, $1/m^2_Q$ corrections and $QCD$ corrections into account. This result agrees with the experimental result $(2.5\sim5.1)\times10^{10}s^{-1}B(\Lambda_c\rightarrow ab)$\cite{DELPHI2004}.\\

\textbf{B. Nonleptonic decays $\Lambda_{b}\rightarrow\Lambda_{c}P(V)$}\\

The Hamiltonian describing nonleptonic decays $\Lambda_b\rightarrow\Lambda_cP(V)$~($P$ and $V$ stand for pesudoscalar and vector mesons, respectively) reads\cite{Buchalla1996}
\begin{eqnarray}
H_{eff}=\frac{G_{F}}{\sqrt{2}}V_{cb}V^{*}_{UD}(a_{1}O_{1}+a_{2}O_{2}),\label{118}
\end{eqnarray}
\noindent with $O_1=(\bar{D}U)(\bar{c}b)$ and $O_2=(\bar{c}U)(\bar{D}b)$, where $U$ and $D$ are the field operators for light quarks involved in the decay, and $(\bar{q}_1q_2)=\bar{q}_1\gamma_{\mu}(1-\gamma_5)q_2$ is understood. The parameters $a_1$ and $a_2$ in Eq. (\ref{118}) are defined as $a_1=c_1+c_2/N_c$, $a_2=c_2+c_1/N_c$, where $c_1$ and $c_2$ are Wilson coefficients, and $N_c$ is the effective color number factor caused by Fiertz transformation. $N_c$ does not equal to 3 in general because of the part that can not be factorized. Therefore, $a_1$ and $a_2$ are treated as free parameters to be determined by experiments. Since $\Lambda_b$ decays are energetic, the factorization assumption is applied so that one of the currents in the Hamiltonian (\ref{118}) is factorized out and generates a meson $P(V)$\cite{Bjorken1989}\cite{Dugan1991}. Thus the decay amplitude of the two body nonleptonic decay becomes the product of two matrix elements, one is related to the decay constant of the factorized meson ($P$ or $V$) and the other is the weak transition matrix element between $\Lambda_b$ and $\Lambda_c$,
\begin{eqnarray}
M^{fac}(\Lambda_{b}\rightarrow\Lambda_{c}P)=\frac{G_{F}}{\sqrt{2}}V_{cb}V^{*}_{UD}<P|A_{\mu}|0><\Lambda_{c}(P')|J^{\mu}|\Lambda_{b}(p)>,\label{119}\\
M^{fac}(\Lambda_{b}\rightarrow\Lambda_{c}P)=\frac{G_{F}}{\sqrt{2}}V_{cb}V^{*}_{UD}<V|V_{\mu}|0><\Lambda_{c}(P')|J^{\mu}|\Lambda_{b}(p)>,\label{120}
\end{eqnarray}
\noindent where $<0|A_{\mu}|P>$ and ~$<0|V_{\mu}|V>$ are related to the decay constants of the pseudoscalar meson or the vector meson respectively by
\begin{eqnarray}
<0|A_{\mu}|P>=if_{P}q_{\mu},\label{121}\\
<0|V_{\mu}|V>=f_{V}m_{V}\epsilon_{\mu},\label{122}
\end{eqnarray}
\noindent where $q_{\mu}$ is the momentum of the meson emitted from the
$W$-boson and $\epsilon_{\mu}$ is the polarization vector of the emitted
vector meson. It is noted that in the two-body nonleptonic
weak decays $\Lambda_b\rightarrow\Lambda_cP(V)$ there is no contribution from the
$a_2$ term since such a term corresponds to the transition of $\Lambda_b$
to a light baryon instead of $\Lambda_c$.
The general form for the amplitude of $\Lambda_b\rightarrow\Lambda_cP(V)$ are
\begin{eqnarray}
M(\Lambda_{b}\rightarrow\Lambda_{c}P)&=&i\bar{u}_{\Lambda_{c}}(P')(A+B\gamma_{5})u_{\Lambda_{b}}(P),\label{123}\\
M(\Lambda_{b}\rightarrow\Lambda_{c}V)&=&\bar{u}_{\Lambda_{c}}(P')\epsilon^{*\mu}(A_{1}\gamma_{\mu}\gamma_{5}+A_{2}P'_{\mu}\gamma_{5}+B_{1}\gamma_{\mu}+B_{2}P'_{\mu})u_{\Lambda_{b}}(P).\label{124}
\end{eqnarray}

On the other hand, the matrix element for $\Lambda_b\rightarrow\Lambda_c$ can be expressed by using Lorentz invariance
\begin{eqnarray}
<\Lambda_{c}(P')|J^{\mu}|\Lambda_{b}(p)>&=&\bar{u}_{\Lambda_{c}}\Big[f_{1}(q^{2})\gamma_{\mu}+if_{2}(q^{2})\sigma_{\mu\nu}q^{\nu}+f_{3}(q^{2})q_{\mu}\nonumber\\
&&-\big(g_{1}(q^{2})\gamma_{\mu}+ig_{2}(q^{2})\sigma_{\mu\nu}q^{\nu}+g_{3}(q^{2})q_{\mu}\big)\gamma_{5}\Big],\label{125}
\end{eqnarray}
\noindent where $f_i$, $g_i~(i=1,2,3)$ are Lorentz scalars. The relations between $f_i$, $g_i$ and form factors $F_i$, $G_i$ in Eq.~(\ref{1}) are
\begin{eqnarray}
f_{1}&=&F_{1}+\frac{1}{2}(m_{\Lambda_{b}}+m_{\Lambda_{c}})\bigg(\frac{F_{2}}{m_{\Lambda_{b}}}+\frac{F_{3}}{m_{\Lambda_{c}}}\bigg),\label{126}\\
f_{2}&=&\frac{1}{2}\bigg(\frac{F_{2}}{m_{\Lambda_{b}}}+\frac{F_{3}}{m_{\Lambda_{c}}}\bigg),\label{127}\\
f_{3}&=&\frac{1}{2}\bigg(\frac{F_{2}}{m_{\Lambda_{b}}}-\frac{F_{3}}{m_{\Lambda_{c}}}\bigg),\label{128}\\
g_{1}&=&G_{1}-\frac{1}{2}(m_{\Lambda_{b}}-m_{\Lambda_{c}})\bigg(\frac{G_{2}}{m_{\Lambda_{b}}}+\frac{G_{3}}{m_{\Lambda_{c}}}\bigg),\label{129}\\
g_{2}&=&\frac{1}{2}\bigg(\frac{G_{2}}{m_{\Lambda_{b}}}+\frac{G_{3}}{m_{\Lambda_{c}}}\bigg),\label{130}\\
g_{3}&=&\frac{1}{2}\bigg(\frac{G_{2}}{m_{\Lambda_{b}}}-\frac{G_{3}}{m_{\Lambda_{c}}}\bigg).\label{131}
\end{eqnarray}

The decay width for $\Lambda_b\rightarrow\Lambda_cP$ is \cite{Cheng1992}\cite{Pakvasa1990}
\begin{eqnarray}
\Gamma(\Lambda_{b}\rightarrow\Lambda_{c}P)=\frac{|\overrightarrow{P}'|}{8\pi}\Bigg[\frac{(m_{\Lambda_{b}}+m_{\Lambda_{c}})^{2}-m^{2}_{P}}{m^{2}_{\Lambda_{b}}}|A|^{2}+\frac{(m_{\Lambda_{b}}-m_{\Lambda_{c}})^{2}-m^{2}_{P}}{m^{2}_{\Lambda_{b}}}|B|^{2}\Bigg],\label{132}
\end{eqnarray}
\noindent where $A$ and $B$ are related to the form factors by
\begin{eqnarray}
A&=&\frac{G_{F}}{\sqrt{2}}V_{cb}V^{*}_{UD}a_{1}f_{P}\big[(m_{\Lambda_{b}}-m_{\Lambda_{c}})f_{1}+m^{2}_{P}f_{3}\big],\label{133}\\
B&=&\frac{G_{F}}{\sqrt{2}}V_{cb}V^{*}_{UD}a_{1}f_{P}\big[(m_{\Lambda_{b}}+m_{\Lambda_{c}})g_{1}-m^{2}_{P}g_{3}\big].\label{134}
\end{eqnarray}

The decay width for $\Lambda_b\rightarrow\Lambda_cV$ is \cite{Cheng1992}\cite{Pakvasa1990}
\begin{eqnarray}
\Gamma(\Lambda_{b}\rightarrow\Lambda_{c}V)=\frac{|\overrightarrow{P}'|}{8\pi}\frac{E_{\Lambda_{c}}+m_{\Lambda_{c}}}{m_{\Lambda_{b}}}\Bigg[2(|S|^{2}+|P_{2}|^{2})+\frac{E^{2}_{V}}{m^{2}_{V}}(|S+D|^{2}+|P_{1}|^{2})\Bigg],\label{135}
\end{eqnarray}
\noindent where
\begin{eqnarray}
S&=&-A_{1},\label{136}\\
D&=&-\frac{|\overrightarrow{P}'|^{2}}{E_{V}(E_{\Lambda_{c}}+m_{\Lambda_{c}})}(A_{1}-m_{\Lambda_{b}}A_{2}),\label{137}\\
P_{1}&=&-\frac{|\overrightarrow{P}'|^{2}}{E_{V}}\Bigg[\frac{m_{\Lambda_{b}}+m_{\Lambda_{c}}}{E_{\Lambda_{c}}+m_{\Lambda_{c}}}B_{1}+m_{\Lambda_{b}}B_{2}\Bigg],\label{138}\\
P_{2}&=&\frac{|\overrightarrow{P}'|}{E_{\Lambda_{c}}+m_{\Lambda_{c}}}B_{1},\label{139}
\end{eqnarray}
with
\begin{eqnarray}
A_{1}&=&-\frac{G_{F}}{\sqrt{2}}V_{cb}V^{*}_{UD}a_{1}f_{V}m_{V}[g_{1}+g_{2}(m_{\Lambda_{b}}-m_{\Lambda_{c}})],\label{140}\\
A_{2}&=&-2\frac{G_{F}}{\sqrt{2}}V_{cb}V^{*}_{UD}a_{1}f_{V}m_{V}g_{2},\label{141}\\
B_{1}&=&\frac{G_{F}}{\sqrt{2}}V_{cb}V^{*}_{UD}a_{1}f_{V}m_{V}[f_{1}-f_{2}(m_{\Lambda_{b}}+m_{\Lambda_{c}})],\label{142}\\
B_{2}&=&2\frac{G_{F}}{\sqrt{2}}V_{cb}V^{*}_{UD}a_{1}f_{V}m_{V}f_{2}.\label{143}
\end{eqnarray}
Then from Eqs. (\ref{126})-(\ref{143}), we obtain the numerical results for $\Lambda_b\rightarrow\Lambda_cP(V)$ decay
widths by using the results for $F_i$ and $G_i~(i=1,2,3)$ at $q^2=m^2_{P,V}$ from BS equations.  In
Tables~\ref{tab:2} and~\ref{tab:3} we list the results for $m_D=0.65GeV$ and $m_D=0.8GeV$, respectively. One can see that the results change only a little with $m_D$. The numbers
without (with) brackets correspond to $\kappa=0.02GeV^3~(
\kappa=0.08GeV^3)$. Again, the subscripts "$0$", "$1/m_Q$",
"$1/m_Q+QCD$", "$1/m_Q+1/m^2_Q$" and "$1/m_Q+1/m^2_Q+QCD$" stand for the results without $1/m_Q$, $1/m^2_Q$ and $QCD$
corrections, with $1/m_Q$ corrections, with both $1/m_Q$ and
$QCD$ corrections, with $1/m_Q$ and $1/m^2_Q$ corrections, and with $1/m_Q$, $1/m^2_Q$ and $QCD$ corrections, respectively. In the calculations we have
taken the following decay constants:
\begin{eqnarray}
&&f_{\pi}=132MeV,~~~~f_{D_{s}}=241MeV,~~~~f_{D}=200MeV,~~~~f_{K}=156MeV.\nonumber\\
&&f_{\rho}=216MeV,~~~~f_{K^{*}}=f_{\rho},~~~~f_D=f_{D^{*}},~~~~f_{D_s}=f_{D^{*}_s}.\nonumber
\end{eqnarray}

\begin{table}[ht!]
\begin{center}
\caption{When $m_{D}=650MeV$, the nonleptonic decay widths for ~$\Lambda_{b}\rightarrow\Lambda_{c}P(V)$ for $\kappa=0.02GeV^3$(~$\kappa=0.08GeV^3$), in units ~$10^{10}s^{-1}a^{2}_{1}$, with $V_{cb}=0.0398/0.042$.}
\begin{tabular}{|c|c c c|}\hline
$Process$ & $\Gamma_{0}$ & $\Gamma_{1/m_{Q}}$ & $\Gamma_{1/m_{Q}+1/m^{2}_{Q}}$  \\ \hline 
$\Lambda^{0}_{b}\rightarrow\Lambda^{+}_{c}\pi^{-}$  & $0.24/0.27(0.45/0.50)$ & $0.28/0.31(0.53/0.58)$ & $0.32/0.35(0.60/0.66)$ \\
\hline $\Lambda^{0}_{b}\rightarrow\Lambda^{+}_{c}D^{-}_{s}$ & $0.77/0.86(1.20/1.33)$ & $0.88/0.98(1.42/1.59)$ & $0.98/1.09(1.57/1.75)$\\
\hline $\Lambda^{0}_{b}\rightarrow\Lambda^{+}_{c}D^{-}$ & $0.028/0.032(0.044/0.049)$ & $0.032/0.036(0.055/0.060)$ & $0.035/0.040(0.062/0.068)$\\
\hline $\Lambda^{0}_{b}\rightarrow\Lambda^{+}_{c}K^{-}$ & $0.017/0.019(0.033/0.036)$ &
$0.020/0.022(0.038/0.043)$ & $0.023/0.026(0.042/0.048)$  \\ 
\hline
$\Lambda^{0}_{b}\rightarrow\Lambda^{+}_{c}\rho^{-}$ &
$0.32/0.35(0.61/0.67)$ & $0.37/0.41(0.70/0.78)$ & $0.43/0.47(0.81/0.89)$ \\
\hline
$\Lambda^{0}_{b}\rightarrow\Lambda^{+}_{c}K^{*-}$ &
$0.017/0.019(0.031/0.035)$ & $0.020/0.022(0.035/0.038)$ & $0.023/0.025(0.038/0.042)$ \\
\hline
$\Lambda^{0}_{b}\rightarrow\Lambda^{+}_{c}D^{*-}$ &
$0.021/0.023(0.032/0.036)$ & $0.023/0.026(0.037/0.041)$ &
$0.025/0.028(0.041/0.046)$ \\
\hline
$\Lambda^{0}_{b}\rightarrow\Lambda^{+}_{c}D^{*-}_{s}$ & $0.59/0.65(0.95/1.04)$ & $0.68/0.76(1.03/1.15)$ & $0.75/0.84(1.11/1.25)$\\
\hline
$Process$ & $\Gamma_{1/m_{Q}+QCD}$ & $\Gamma_{1/m_{Q}+1/m^{2}_{Q}+QCD}$ & \\
\hline $\Lambda^{0}_{b}\rightarrow\Lambda^{+}_{c}\pi^{-}$ & $0.22/0.25(0.43/0.48)$ & $0.26/0.30(0.53/0.59)$ & \\
\hline $\Lambda^{0}_{b}\rightarrow\Lambda^{+}_{c}D^{-}_{s}$ &
$0.76/0.85(1.21/1.34)$ & $0.90/0.98(1.31/1.46)$\\
\hline $\Lambda^{0}_{b}\rightarrow\Lambda^{+}_{c}D^{-}$ & $0.027/0.030(0.043/0.048)$ & $0.030/0.034(0.048/0.055)$ &\\
\hline $\Lambda^{0}_{b}\rightarrow\Lambda^{+}_{c}K^{-}$ & $0.016/0.018(0.032/0.035)$ & $0.020/0.022(0.037/0.042)$ &\\
\hline
$\Lambda^{0}_{b}\rightarrow\Lambda^{+}_{c}\rho^{-}$ &
$0.30/0.33(0.58/0.65)$ & $0.34/0.38(0.64/0.0.72)$ &\\
\hline
$\Lambda^{0}_{b}\rightarrow\Lambda^{+}_{c}K^{*-}$ &
$0.016/0.018(0.031/0.034)$ & $0.018/0.020(0.033/0.036)$ &\\
\hline
$\Lambda^{0}_{b}\rightarrow\Lambda^{+}_{c}D^{*-}$ &
$0.020/0.022(0.031/0.035)$ & $0.022/0.024(0.036/0.042)$ &\\
\hline
$\Lambda^{0}_{b}\rightarrow\Lambda^{+}_{c}D^{*-}_{s}$ & $0.57/0.64(0.94/1.03)$ & $0.63/0.71(1.01/1.12)$ &\\
\hline
\end{tabular}\label{tab:2}
\end{center}
\end{table}

\begin{table}[ht!]
\begin{center}
\caption{When $m_{D}=800MeV$, the nonleptonic decay widths for ~$\Lambda_{b}\rightarrow\Lambda_{c}P(V)$ for $\kappa=0.02GeV^3$(~$\kappa=0.08GeV^3$), in units ~$10^{10}s^{-1}a^{2}_{1}$, with $V_{cb}=0.0398/0.042$.}
\begin{tabular}{|c|c c c|}\hline
$Process$ & $\Gamma_{0}$ & $\Gamma_{1/m_{Q}}$ & $\Gamma_{1/m_{Q}+1/m^{2}_{Q}}$  \\ \hline 
$\Lambda^{0}_{b}\rightarrow\Lambda^{+}_{c}\pi^{-}$  & $0.27/0.30(0.50/0.55)$ & $0.31/0.35(0.57/0.64)$ & $0.35/0.39(0.66/0.73)$ \\
\hline $\Lambda^{0}_{b}\rightarrow\Lambda^{+}_{c}D^{-}_{s}$ & $0.84/0.93(1.28/1.41)$ & $0.96/1.06(1.53/1.74)$ & $1.05/1.17(1.72/1.92)$\\
\hline $\Lambda^{0}_{b}\rightarrow\Lambda^{+}_{c}D^{-}$ & $0.031/0.035(0.048/0.054)$ & $0.036/0.040(0.060/0.066)$ & $0.039/0.044(0.068/0.074)$\\
\hline $\Lambda^{0}_{b}\rightarrow\Lambda^{+}_{c}K^{-}$ & $0.020/0.022(0.037/0.040)$ &
$0.023/0.025(0.042/0.046)$ & $0.026/0.029(0.046/0.053)$  \\ 
\hline
$\Lambda^{0}_{b}\rightarrow\Lambda^{+}_{c}\rho^{-}$ &
$0.36/0.40(0.66/0.73)$ & $0.41/0.46(0.77/0.85)$ & $0.45/0.52(0.86/0.94)$ \\
\hline
$\Lambda^{0}_{b}\rightarrow\Lambda^{+}_{c}K^{*-}$ &
$0.019/0.022(0.035/0.040)$ & $0.022/0.025(0.041/0.046)$ & $0.024/0.028(0.046/0.052)$ \\
\hline
$\Lambda^{0}_{b}\rightarrow\Lambda^{+}_{c}D^{*-}$ &
$0.023/0.026(0.036/0.039)$ & $0.026/0.029(0.040/0.045)$ & $0.029/0.032(0.044/0.050)$ \\
\hline
$\Lambda^{0}_{b}\rightarrow\Lambda^{+}_{c}D^{*-}_{s}$ & $0.65/0.73(1.04/1.15)$ & $0.75/0.84(1.15/1.27)$ & $0.84/0.94(1.24/1.38)$ \\
\hline
$Process$ & $\Gamma_{1/m_{Q}+QCD}$ & $\Gamma_{1/m_{Q}+1/m^{2}_{Q}+QCD}$ & \\
\hline $\Lambda^{0}_{b}\rightarrow\Lambda^{+}_{c}\pi^{-}$ & $0.25/0.28(0.48/0.53)$ & $0.29/0.34(0.55/0.63)$ & \\
\hline $\Lambda^{0}_{b}\rightarrow\Lambda^{+}_{c}D^{-}_{s}$ &
$0.82/0.91(1.27-1.40)$ & $0.95/1.04(1.35/1.51)$ &\\
\hline $\Lambda^{0}_{b}\rightarrow\Lambda^{+}_{c}D^{-}$ & $0.030/0.034(0.047/0.053)$ & $0.034/0.039(0.051/0.058)$ &\\
\hline $\Lambda^{0}_{b}\rightarrow\Lambda^{+}_{c}K^{-}$ & $0.019/0.021(0.035/0.039)$ & $0.022/0.025(0.042/0.047)$ & \\
\hline
$\Lambda^{0}_{b}\rightarrow\Lambda^{+}_{c}\rho^{-}$ &
$0.34/0.38(0.64/0.71)$ & $0.38/0.44(0.72/0.89)$ &\\
\hline
$\Lambda^{0}_{b}\rightarrow\Lambda^{+}_{c}K^{*-}$ &
$0.018/0.020(0.034/0.038)$ & $0.020/0.023(0.038/0.042)$ & \\
\hline
$\Lambda^{0}_{b}\rightarrow\Lambda^{+}_{c}D^{*-}$ &
$0.022/0.025(0.034/0.037)$ & $0.024/0.028(0.037/0.041)$ & \\
\hline
$\Lambda^{0}_{b}\rightarrow\Lambda^{+}_{c}D^{*-}_{s}$ & $0.63/0.70(1.02/1.14)$ & $0.71/0.79(1.13/1.25)$ & \\
\hline
\end{tabular}\label{tab:3}
\end{center}
\end{table}

From Tables~\ref{tab:2} and~\ref{tab:3}, one can see that $1/m_Q$ and $1/m^2_Q$ corrections enlarge the decay widths while $QCD$ corrections reduce them. Taking $\Lambda^{0}_{b}\rightarrow\Lambda^{+}_{c}\pi^{-}$ as an example, the decay width with $1/m_Q$ corrections is $17\%$ bigger than that without any corrections, and the decay width with $1/m_Q$ and $1/m_Q^2$ corrections is $14\%$ larger than that with just $1/m_Q$ correction. The experimental data for decay width of $\Lambda_b\rightarrow\Lambda_c\pi$ is $(0.14\sim0.69)\times10^{10}s^{-1}$\cite{CDF2007122002}, and the numerical result from our model is $(0.26\sim0.63)\times10^{10}s^{-1}a^2_1$. As we know, $a_1=c_1+c_2/N_c$, $c_1=1.1502$, $c_2=-0.3125$\cite{Deshpande1995}\cite{Fleischer1997}, $N_c=2\sim5$\cite{Guo20072007}\cite{Chen1999}, so that $a_1\sim O(1)$. Our theoretical result matches the experimental data if we take $a\simeq1$. There are not experimental data for other processes. Our predictions for these processes will be tested in the future.

\section*{V. Summary and discussion}

In the present work, we assume that a heavy baryon $\Lambda_Q$ is composed of a heavy quark and a scalar light diquark. Besed on this picture we analyze the $1/m^2_Q$ corrections to the BS equation for $\Lambda_Q$ and apply the results to calculate the six weak decay form factors $F_i$, $G_i~(i=1,2,3)$. We find that they mostly depend on $\kappa$ and depend only a little on the diquark mass, $m_D$. Then we apply the numerical results of these six form factors  to calculate the differential and total decay widths for the semileptonic decay $\Lambda_b\rightarrow\Lambda_cl\bar{\nu}$, and the nonleptonic decay widths for $\Lambda_b\rightarrow\Lambda_cP(V)$. Not only $1/m_Q$ and $1/m_Q^2$ corrections, but also $QCD$ corrections are taken into account because the last one is comparable with $1/m_Q$ corrections. One can see that the numerical results for the decay widths mostly depend on $\kappa$. The decay width for $\Lambda_b\rightarrow\Lambda_cl\bar{\nu}$ is ($2.15\sim3.60)\times10^{10}s^{-1}$ when we consider all the corrections mentioned above, and it matches the experimental data, ($2.5\sim5.1)\times10^{10}s^{-1}$. The decay width for $\Lambda_b\rightarrow\Lambda_c\pi$ is $(0.26\sim0.63)\times10^{10}s^{-1}a^2_1$, and it agrees with the experimental data, $(0.14\sim0.69)\times10^{10}s^{-1}$,  when the color factor $N_c=2\sim5$, and hence, $a_1\simeq1$. We also give predictions for other processes including $\Lambda_b\rightarrow\Lambda_cD_s^-$, $\Lambda_b\rightarrow\Lambda_cD^-$, $\Lambda_b\rightarrow\Lambda_cK^-$, $\Lambda_b\rightarrow\Lambda_cP^-$, $\Lambda_b\rightarrow\Lambda_cK^{\star-}$, $\Lambda_b\rightarrow\Lambda_cD^{\star-}$, and $\Lambda_b\rightarrow\Lambda_cD_s^{\star-}$. These predictions will be tested in future experiments.



\begin{thebibliography}{s4}

\bibitem{OPAL}  OPAL Collaboration, R. Akers et al., Z. Phys. C {\bf69}, 195
(1996); Phys. Lett. B {\bf353}, 402 (1995); OPAL Collaboration,
K. Ackerstaff et al., \textit{ibid}. {\bf 426}, 161 (1998).

\bibitem{PDG14}   Particle Data Group, July 2014 Particle Physics Booklet (2014).

\bibitem{Isgur1989} N. Isgur and M.B. Wise, Phys. Lett. B {\bf232}, 113 (1989); {\bf237}, 527 (1990); H. Georgi, \textit{ibid}. {\bf264}, 447 (1991); see also M. Neubert, Phys. Rep. {\bf245}, 259 (1994) for the review.

\bibitem{Grin1990}
B. Grinstein, Nucl. Phys. B \textbf{339} (1990) 252.

\bibitem{Voloshin1987}
M. B. Voloshin and M. A. Shifman, Sov. J. Nucl. Phys. \textbf{45}, (1987) 292; $ibid$. \textbf{47}, (1988) 511; H.
Politzer and M. B. Wise, Phys. Lett. B \textbf{206}, (1988) 681; $ibid$. B \textbf{208}, (1988) 504; E. Eichten and B. Hill, Phys. Lett. B \textbf{234} (1990) 511; J.D. Bjorken in Results and Perspectives in Particle Physics, proc. of the 4th Rencontres de Physique de la Valle $d^{,}$ Aoste, La Thuile, Italy 1990, M. Greco ed. (ed. Frontieres, Gif-sur-Yvette, France, 1990) p. 583; N. Isgur and M. B. Wise, Phys. Rev. D \textbf{43}, (1991) 819.

\bibitem{Falk1990}
A. Falk, H. Georgi, B. Grinstein and M. B. Wise, Nucl. Phys. B \textbf{343}, (1990) 1.

\bibitem{Georgi1990}
H. Georgi, Phys. Lett. B \textbf{240}, (1990) 447.

\bibitem{Akers1996}
OPAL Collaboration, R. Akers $et~~al.$, Z. Phys. C \textbf{69}
(1996) 195; Phys. Lett. B. \textbf{353} (1995) 402; OPAL
Collaboration, K. Ackerstaff $et~~al.$, $ibid$. \textbf{426} (1998)
161.

\bibitem{Albarjar1991}
UAl Collaboration, C. Albarjar $et~~al.$, Phys. Lett. B \textbf{273}
(1991) 540.

\bibitem{Abe1993}
CDF Collaboration, F. Abe $et~~al.$, Phys. Rev. D \textbf{47} (1993)
2639.

\bibitem{Tzmarias1994}
S. E. Tzmarias, invited talk presented in the 27th International
Conference on High Energy Physics, Glasgow, 1994; P. Abreu
$et~~al.$, Phys. Lett. B \textbf{374} (1996) 351.

\bibitem{Abe1997}
CDF Collaboration, F. Abe $et~~al.$, Phys. Rev. D \textbf{55} (1997)
1142.

\bibitem{CDF1997}
CDF Collaboration, F. Abe $et al.$, Phys. Rev. D \text{55}, (1997) 1142; UA1 Collab-
oration, C. Albarjar $et al.$, Phys. Lett. B \textbf{273}, (1991) 540.

\bibitem{DELPHI1995}
DELPHI Collaboration, P. Abreu $et al.$, Z. Phys. C \textbf{68}, (1995) 375; ALEPH
Collaboration, R. Barate $et al.$, Eur. Phys. J. C \textbf{2}, (1998) 197.

\bibitem{DELPHI2004}
DELPHI Collaboration, J. Abdallah $et al.$, Phys. Lett. B \textbf{585}, (2004) 63.

\bibitem{CDF2007122002}
CDF Collaboration, A. Abulencia $et al.$, Phys. Rev. Lett. \textbf{98}, (2007) 122002.

\bibitem{CDF2011012003}
CDF Collaboration, T. Aaltonen $et al.$, Phys. Rev.
\textbf{D84} (2011) 012003; CLEO Collaboration, S. B. Athar $et al.$, Phys. Rev. \textbf{D71} (2005) 051101;
CLEO Collaboration, R. Ammar $et al.$, Phys. Rev. Lett \textbf{86} (2001) 1167;BarBar Collaboration, B. Aubert $et al.$, Phys. Rev. \textbf{D78} (2008) 112003; CLEO Collaboration, M. Artuso $et al.$, Phys. Rev. \textbf{D65} (2002) 071101.

\bibitem{CDF2007202001}
CDF Collaboration, T. Aaltonen $et al.$, Phys. Rev. Lett. \textbf{99}, (2007) 202001.

\bibitem{Georgi1990}
H. Georgi, B. Grinstein, and M.B. Wise, Phys. Lett. B {\bf252}, 456 (1990).

\bibitem{Guo1996}
X.-H. Guo and T. Muta, Phys. Rev. D {\bf54}, 4629 (1996); Mod.
Phys. Lett. A {\bf11}, 1523 (1996).

\bibitem{Lurie1968}
D. Lurie, \textit{Particles and Fields} (Interscience Publishers, John
Wiley $\&$ Sons, New York, London, Sydney, 1968); C. Itzyk-
son and J.B. Zuber, \textit{Quantum Field Theory} (McGraw-Hill,
New York, 1980).

\bibitem{Zhang2013}
L. Zhang and X.-H. Guo, Phys. Rev. D {\bf87}, 076013 (2013).

\bibitem{Ansel1987}
M. Anselmino, P. Kroll, and B. Pire, Z. Phys. {\bf C 36}, 89 (1987)

\bibitem{Jin1992}
H.-Y. Jin, C.-S. Huang, and Y.-B. Dai, Z. Phys. C {\bf56}, 707
(1992); Y.-B. Dai, C.-S. Huang, and H.-Y. Jin, \textit{ibid}. {\bf60}, 527
(1993); {\bf65}, 87 (1995).

\bibitem{Korner1992}
J.G. K$\ddot{o}$rner and M. Kr$\ddot{a}$mer, Phys. Lett. B \textbf{275}, (1992) 495.

\bibitem{Neubert1992}
 M. Neubert, Nucl. Phys. \textbf{B371}, 149 (1992)

\bibitem{Buchalla1996}
G. Buchalla, A. J. Buras, and M. E. Lautenbacher, Rev. Mod. Phys. \textbf{68}, (1996) 1125.

\bibitem{Bjorken1989}
J.D. Bjorken, Nucl. Phys. B (Proc. Suppl.) \textbf{11}, (1989) 325.

\bibitem{Dugan1991}
M.J. Dugan and B. Grinstein, Phys. Lett. B \textbf{255}, (1991) 583.

\bibitem{Cheng1992}
H.Y. Cheng, Phys. Lett. B \textbf{289}, (1992) 455; Phys. Rev. \textbf{D 56}, (1997) 2799; H.Y. Cheng and B. Tseng, $ibid.$ \textbf{53}, (1996) 1457.

\bibitem{Pakvasa1990}
S. Pakvasa, S.F. Tuan, and S.P. Rosen, Phys. Rev. \textbf{D 42}, (1990) 3746.

\bibitem{Deshpande1995}
N. G. Deshpande and X.-G. He, Phys. Rev. Lett. \textbf{74}, (1995) 26.

\bibitem{Fleischer1997}
R. Fleischer, Int. J. Mod. Phys. A \textbf{12}, (1997) 2459; Z. Phy. C \textbf{62}, (1994) 81; \textbf{58}, (1993) 483.

\bibitem{Guo20072007}
X.-H. Guo and Z.-H. Zhang, Phys. Rev. \textbf{D 75}, (2007) 074028.

\bibitem{Chen1999}
Y.-H. Chen, H.-Y. Cheng, B. Tseng, and K.-C. Yang, Phys. Rev. \textbf{D 60}, (1999) 094014.






\end{thebibliography}
\end{document}